\documentclass[twocolumn,natbib]{svjour3}         
\smartqed  
\usepackage{graphicx}
\newcommand{\aap}{A\&A\ }

\newcommand{\apj}{ApJ\ }

\newcommand{\mnras}{MNRAS\ }

\def \xmm {{\em XMM-Newton}}

\def \hcm {\hbox {\ifmmode $ atom cm$^{-2}\else atom cm$^{-2}$\fi}}

\def \arcsec {\hbox{$^{\prime\prime}$}}

\def\approxgt{\mathrel{\hbox{\rlap{\lower.55ex \hbox {$\sim$}}
        \kern-.3em \raise.4ex \hbox{$>$}}}}
\def\approxlt{\mathrel{\hbox{\rlap{\lower.55ex \hbox {$\sim$}}
        \kern-.3em \raise.4ex \hbox{$<$}}}}

\newcommand\phib{\phi_{\bf B}}
\newcommand\phigb{\phi_{\nabla |{\bf B}|}}
\newcommand\gb{\nabla |{\bf B}|}

\def\lsim{\;\raise0.3ex\hbox{$<$\kern-0.75em\raise-1.1ex\hbox{$\sim$}}\;}
\def\gsim{\;\raise0.3ex\hbox{$>$\kern-0.75em\raise-1.1ex\hbox{$\sim$}}\;}

\def\beq{\begin{equation}}
\def\enq{\end{equation}}
\def\begar{\begin{eqnarray}}
\def\endar{\end{eqnarray}}
\def\mathnew{\mathsurround=0pt}
\def\simov#1#2{\lower .5pt\vbox{\baselineskip0pt \lineskip-.5pt
        \ialign{$\mathnew#1\hfil##\hfil$\crcr#2\crcr\sim\crcr}}}

\def \xmm {{\it XMM-Newton}}

\catcode`\@=11
\newcommand{\gapprox}{\mathrel{\mathpalette\@versim>}}
\newcommand{\lapprox}{\mathrel{\mathpalette\@versim<}}
\newcommand{\propapprox}{\mathrel{\mathpalette\@versim\propto}}
\newcommand{\@versim}[2]
  {\lower3.1truept\vbox{\baselineskip0pt\lineskip0.5truept
\ialign{$\m@th#1\hfil##\hfil$\crcr#2\crcr\sim\crcr}}}
\catcode`\@=12

\begin{document}

\title{Magnetic fields in supernova remnants and pulsar-wind nebulae
}


\author{Stephen P. Reynolds         \and
        B. M. Gaensler \and
        Fabrizio Bocchino 
}


\institute{S.P. Reynolds \at
              Physics Department, North Carolina State University \\
	      Raleigh, NC, USA\\
              Tel.: +01-919-515-7751\\
              Fax: +01-919-515-6538\\
              \email{reynolds@ncsu.edu}           
           \and
            B.M. Gaensler \at
         Sydney Institute for Astronomy \\
         School of Physics, The University of Sydney \\
         NSW 2006, Australia \\          
         \email{bryan.gaensler@sydney.edu.au} 
           \and
           Fabrizio Bocchino \at
           INAF-Osservatorio Astronomico di Palermo \\
	   Piazza del Parlamento 1 \\
	   90134 Palermo, Italy \\
	   \email{bocchino@astropa.inaf.it}
}

\date{Received: date / Accepted: date}

\maketitle

\begin{abstract}
We review the observations of supernova remnants (SNRs) and
pulsar-wind nebulae (PWNe) that give information on the strength and
orientation of magnetic fields.  Radio polarimetry gives the degree of
order of magnetic fields, and the orientation of the ordered
component.  Many young shell supernova remnants show evidence for
synchrotron X-ray emission.  The spatial analysis of this emission
suggests that magnetic fields are amplified by one to two orders of
magnitude in strong shocks.  Detection of several remnants in TeV
gamma rays implies a lower limit on the magnetic-field strength (or a
measurement, if the emission process is inverse-Compton upscattering
of cosmic microwave background photons).  Upper limits to GeV emission
similarly provide lower limits on magnetic-field strengths.  In the
historical shell remnants, lower limits on $B$ range from 25 to 1000
$\mu$G.  Two remnants show variability of synchrotron X-ray emission
with a timescale of years.  If this timescale is the
electron-acceleration or radiative loss timescale, magnetic fields of
order 1 mG are also implied.  In pulsar-wind nebulae, equipartition
arguments and dynamical modeling can be used to infer magnetic-field
strengths anywhere from $\sim 5\ \mu$G to 1 mG.  Polarized fractions
are considerably higher than in SNRs, ranging to 50 or 60\% in some
cases; magnetic-field geometries often suggest a toroidal structure
around the pulsar, but this is not universal.  Viewing-angle effects
undoubtedly play a role.  MHD models of radio emission in shell SNRs
show that different orientations of upstream magnetic field, and
different assumptions about electron acceleration, predict different
radio morphology.  In the remnant of SN 1006, such comparisons imply a
magnetic-field orientation connecting the bright limbs, with a
non-negligible gradient of its strength across the remnant.

\end{abstract}

\section{Shell supernova remnants: review}

\subsection{Introduction}
\label{intro}

Supernova remnants and pulsar-wind nebulae are prominent Galactic
synchrotron sources at radio and, often, X-ray wavelengths.  The
spatial and spectral analysis of the synchrotron emission can be used
to deduce or constrain magnetic-field strengths and orientations.
In this review, ``supernova remnant'' (SNR) will be used for shell
(i.e., non-pulsar-driven) remnants; if the emission is due to a pulsar,
whether the object is young (like the Crab Nebula) or old (as in
H$\alpha$ bow-shock nebulae) the object will be termed a pulsar-wind
nebula (PWN).  Thus a SNR may contain a PWN, a combination sometimes
called a ``composite'' SNR.  

SNRs are primarily radio objects.  In the Milky Way, 274 are
listed in Green's (2009) well-known catalogue 
(http://www.mrao.cam.ac.uk/surveys/snrs/); most are known
only by radio emission.  SNRs are well-studied only in Local Group
galaxies where they are bright and large enough to be well-imaged with
radio interferometers.  Galactic remnants range in angular size from
less than $2'$ to many degrees, and represent a range in ages from
about 100 years (G1.9+0.3) to over $10^5$ yr, though age estimates
are difficult for very old objects.

\subsection{SNR Dynamics}

SNRs evolve through various phases as their shock waves decelerate in
the ambient medium.  Very soon after an explosion (either Type Ia or
core-collapse) deposits $\sim 10^{51}$ erg into the ISM, a shock wave
is driven into the surrounding circumstellar medium (CSM: modified by
the progenitor) or unmodified interstellar medium (ISM).
Core-collapse SNe eject several solar masses with a range of
velocities, typically of order 5,000 km s$^{-1}$.  Type Ia's are
thought to represent the thermonuclear disruption of a white dwarf, so
eject 1.4 $M_\odot$ at about 10,000 km s$^{-1}$.  Not long after the
explosion, deceleration of the outer blast wave causes the formation
of a reverse shock as inner ejecta are forced to decelerate.  This
begins the ejecta phase of evolution, characterized by the presence of
both the forward shock (blast wave) heating ISM or CSM, and the
reverse shock heating ejecta.  Cooling timescales are much longer than
dynamical timescales, so this phase is essentially adiabatic.
Depending on the density structure in the ejecta and surrounding
material, the shock radius may evolve as $t^{0.6} - t^{0.9}$
\citep{chevalier82}.  Eventually the reverse shock moves in to the
center of the SNR and, after what may be extensive reverberations,
disappears.  By this time (several thousand years for typical
parameters), several times the ejected mass have been swept up by the
blast wave, and the remnant is settling into the Sedov phase, well
described by the similarity solution for a point explosion in a
uniform medium, with shock radius obeying $R \propto t^{0.4}$.  This
phase is adiabatic as well.  Eventually, cooling times in shocked ISM
become comparable to the remnant age, and shocks become radiative:
lossy and optically prominent.  This typically occurs when shock
velocities $v_s$ drop to around 200 km s$^{-1}$.  Denser regions
around the remnant periphery may become radiative before other parts;
even young SNRs (like Kepler, SN 1604) can show radiative
shocks in some regions.  Once the bulk of the blast wave is radiative,
the deceleration is more marked.  The remnant interior will remain hot
for some time, producing what is known as a ``pressure-driven
snowplow.''  Eventually, the remnant becomes sufficiently confused as
to lose its identity, and the remaining kinetic energy is dissipated
as sound waves in the ISM.  Various blast-wave solutions are described
in \cite{ostriker88}; particular applications to SNRs are made in
\cite{truelove99}.

\subsection{Radio inferences}

\begin{figure}
\includegraphics[width=\hsize]{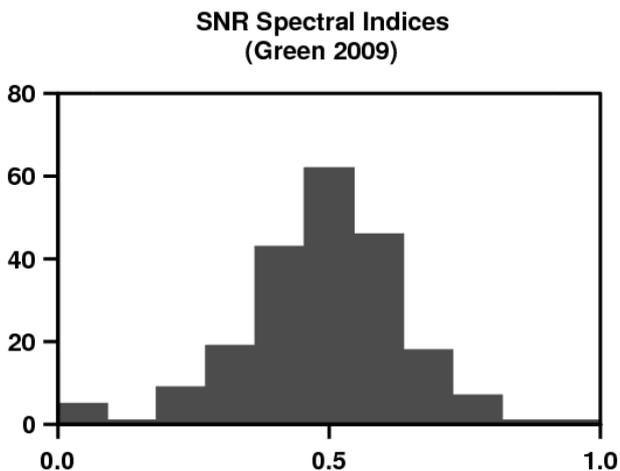}
\caption{Histogram of shell SNRs with fairly well-measured radio
spectral indices, from Green 2009.  PWNe are excluded.}
\label{spix}
\end{figure}

The prevalence of radio emission from adiabatic-phase SNRs, with low
shock compression ratios, indicates that particle acceleration must be
ongoing; simple compression of ambient Galactic magnetic field and
cosmic-ray electrons would produce both insufficiently bright emission
and the wrong spectrum \citep{reynolds08a}.  Since 1977, the
traditional explanation for this electron acceleration has been
diffusive shock acceleration (DSA; see Blandford \& Eichler 1987 for a
comprehensive review).  Electrons emitting synchrotron radiation at
radio wavelengths have energies of order 1 -- 10 GeV ($E = 14.7
\sqrt{\nu({\rm GHz})/B(\mu{\rm G})}$ GeV).  The observed radio
power-law spectra $S_\nu \propto \nu^{-\alpha}$, with $\alpha \sim
0.5$ (see Figure~\ref{spix}), imply from synchrotron theory a
power-law energy distribution of electrons $KE^{-s}$ electrons
cm$^{-3}$ erg$^{-1}$ with $s = 2\alpha + 1 \sim 2$.  Radio
observations do not allow the separate deduction of energy in
electrons and in magnetic field; the synchrotron emissivity of this
power-law distribution of electrons is proportional to $KB^{(s+1)/2}$,
or roughly to the product of energy density in electrons and in
magnetic field.  If particle and magnetic-field energies were in
equipartition (the minimum-energy state producing an observed
spectrum), the magnetic-field strength could be deduced.  For a
spherical remnant of radius $R$ pc at distance $D$ kpc with $\alpha =
0.5$,
\begin{equation}
B = 20 \left( (1 + k) S_9 D_{\rm kpc}^2/(\phi R_{\rm pc}^3)\right)^{2/7}
\ \mu{\rm G},
\end{equation}
where $S_9$ is the flux density at 1 GHz in Jy, $\phi$ is the volume
filling factor, and $k$ is the ratio of energy density in ions to that
in electrons (e.g., Pacholczyk 1970).  But there is no obvious reason
that equipartition should hold: SNRs are typically very inefficient
synchrotron radiators, with a very small fraction of the $10^{51}$ erg
required to produce the observed emission.  Thermal energies are very
much larger than either particle or field energy.  Thus equipartition
magnetic-field strengths are really just a measure of (approximately)
remnant surface brightness.

\begin{figure}
\includegraphics[width=\hsize]{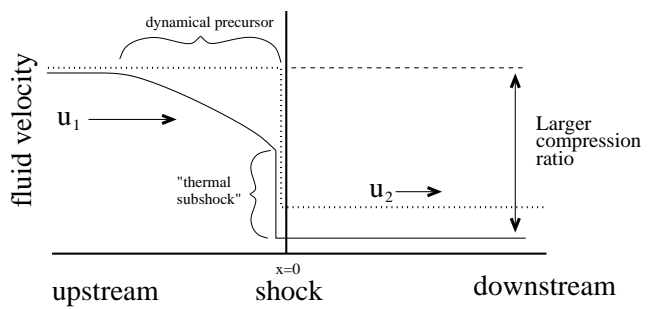}
\caption{Schematic velocity profile of a shock wave with upstream
velocity $u_1$ and downstream $u_2$.  Dotted line:  test-particle
shock (velocity discontinuity).  Solid line:  shock modified by
cosmic rays.}
\label{shockfig}
\end{figure}

The question of magnetic-field amplification in strong SNR shocks, to
be discussed at length below, is intimately related to the question of
shock modification by cosmic-ray pressure (efficient shock
acceleration, or nonlinear DSA).  Thus evidence for nonlinear DSA is
related to estimation of magnetic-field strengths.  One prediction of
this theory is the gradual deceleration of incoming flow (in the shock
frame) by forward-diffusing cosmic rays, so that the compression ratio
$r$ varies continuously from a possibly large value (far upstream
velocity/downstream velocity, perhaps 10 or more) to about 3, at which
point most calculations show the development of a ``thermal subshock''
on a length scale of thermal proton gyroradius, much smaller than
other length scales.  (See Figure~\ref{shockfig}.) In test-particle
(inefficient) DSA, well-known results give the spectrum of accelerated
particles as a power-law with energy index $s$ depending only on the
shock compression ratio, $s = (r + 2)/(r - 1)$.  Since strong
adiabatic shocks are expected to have $r = 4$ (for adiabatic index
$\gamma = 5/3$), we predict $s = 2$ which implies $\alpha = 0.5$, close
to the mean of the distribution of over 200 SNR values from Green's
catalogue.  However, the dispersion in that distribution is
substantial (see Figure~\ref{spix}); explaining values of $\alpha$ of
0.6 -- 0.7, typical for young SNRs, or as low as 0.3, sometimes seen
in older SNRs, is not straightforward.  Low Mach-number shocks will
have lower compressions: in fact, for Mach number $\cal M$, $r^{-1} =
(\gamma - 1)/(\gamma + 1) + 2/((\gamma + 1){\cal M}^2)$.  For ${\cal
M} < 10$, this can significantly increase the steepness of the
predicted spectrum; but this is demonstrably not the case in the young
remnants with the steeper spectra.  No variation in shock Mach number
can explain $\alpha < 0.5$; traditional explanations involve confusing
thermal emission or other instrumental effects, but remain unconvincing.

A consistent explanation for steeper radio spectra in young remnants
is available, however.  Particles of a particular energy will diffuse
(on average) a certain distance ahead of an efficient shock, where
they will see a particular effective compression ratio, which fixes
the slope of the distribution function near that energy.  If, as
expected, more energetic particles diffuse further, they will see
larger compression ratios and the spectrum will become flatter.  Thus
one predicts spectra flattening (hardening) to higher energy, an
effect first pointed out by Eichler (1979).  This effect was
quantified in a Monte Carlo simulation of electron acceleration, and
compared with radio observations of young SNRs, which did seem to show
the effect (Ellison \& Reynolds 1991; Reynolds \& Ellison 1992; see
Figure 3).  The prediction for the emitted synchrotron spectrum
depends on the magnetic-field strength; stronger fields mean
lower-energy particles producing radiation in a particular
radio-astronomical bandpass.  Reynolds \& Ellison found that
explaining the relatively steep spectra of Tycho's and Kepler's SNRs
required magnetic fields of up to 1 mG and above, though with
substantial errors.

\begin{figure}
\includegraphics[width=\hsize]{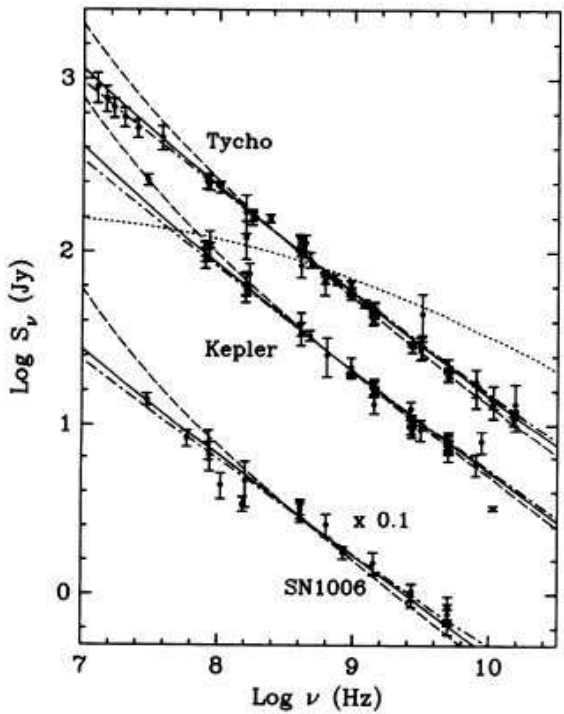}
\caption{Observed total flux observations of Tycho's and Kepler's
SNRs and SN 1006, with model spectra superposed.  All three
spectra derive from the same electron spectrum but with varied
magnetic field.  For Tycho and Kepler, the dashed line corresponds
to $B = 10$ mG, the solid line to 1 mG, and the dot-dashed line
to 0.1 mG; for SN 1006, the dashed, solid, and dot-dashed lines
correspond to $B = 10$ mG, 0.1 mG, and 1 $\mu$G, respectively.
The dotted line shows the spectrum that would be produced if electrons
had the same energy distribution as protons, showing how differently
the nonlinear shock treats electrons and protons.
(Reynolds \& Ellison 1992)}
\end{figure}

\begin{figure}
\includegraphics[width=\hsize]{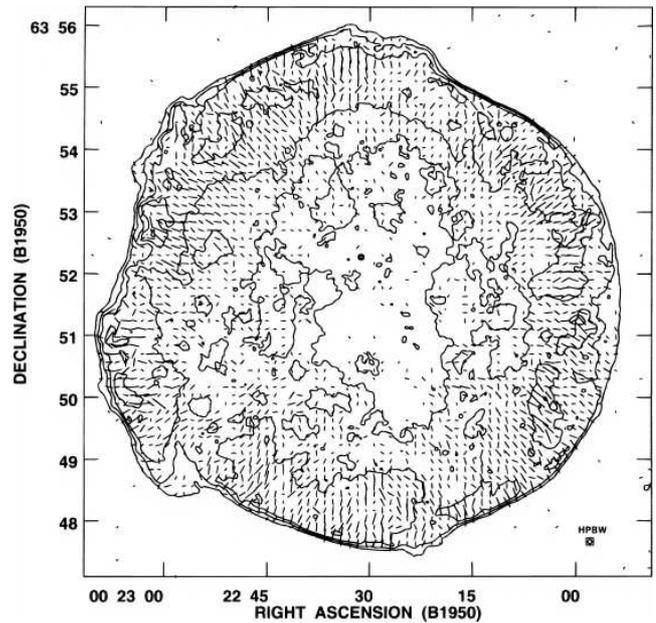}
\caption{Radio contours and polarized $E$-vectors at 1420 MHz for DA
530 (Landecker et al.~1999).  Direction of the sky-plane component of
the magnetic field is at right angles to the directions of the
$E$-vectors, i.e., predominantly tangential in the bright limbs.
Foreground Faraday rotation has been corrected for, though it is
small.}
\label{da530}
\end{figure}

\begin{figure}
\includegraphics[width=\hsize]{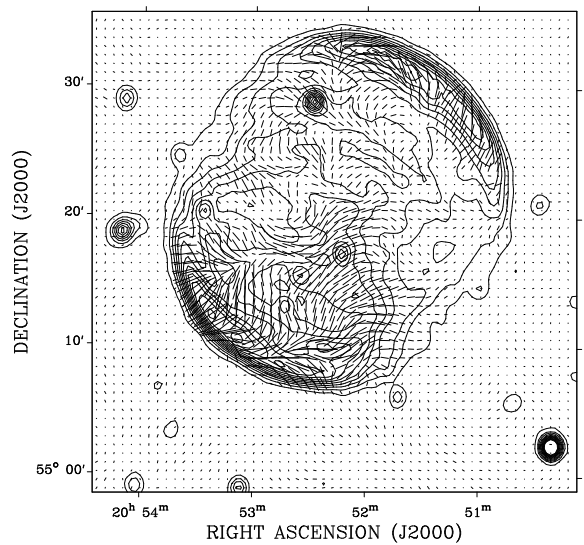}
\caption{Polarization vectors in Tycho's SNR at 1375 MHz (Reynoso et
al.~1997).  These have been corrected for foreground Faraday rotation
and rotated by $90^\circ$ to show the direction of the magnetic field.}
\label{tychopol}
\end{figure}

The degree of order in the magnetic field in SNRs can be directly
inferred from the degree of polarization of synchrotron radiation
(subject to observational issues such as Faraday depolarization and
resolution effects [beam depolarization]).  The degree of linear
polarization of power-law synchrotron radiation from electrons in a
uniform magnetic field is given by $P = (s + 1)/(s + 7/3)$, so that $s
= 2 \Rightarrow P = 69$\%.  Such high values are never observed in
SNRs; a few older SNRs such as DA 530 show $P \sim 50$\% in some
regions \citep[][Figure~\ref{da530}]{landecker99}, but historical
shells show much lower values, 10 -- 15\% (see references in Reynolds
\& Gilmore 1993).  Very roughly, the observed polarized fraction over
the maximum possible gives an estimate of the fractional energy in the
ordered component of the magnetic field \citep{burn66}; thus in Tycho,
for instance (see Figure~\ref{tychopol} [Reynoso et al.~1997]), less
than 15\% of the magnetic energy is in an ordered component.  However,
the direction of that component is radial, a peculiar property shared
by the other historical shells SN 1006, Cas A, Kepler, and G11.2-0.3.
The nature of this radial component is unclear; while radial motions
such as predicted to occur at the Rayleigh-Taylor unstable contact
discontinuity between shocked ejecta and shocked ISM might enhance
radial field, this effect should occur somewhat interior, while radial
fields are observed immediately at remnant edges (e.g., Tycho: Dickel
et al. 1991).  Various MHD simulations have been performed in an
attempt to understand this effect \citep{jun96,jun99}.  In older
remnants such as DA 530, magnetic fields are generally confused or
tangential.  Tangential fields can readily be explained by simple
compression in radiative shocks with large $r$, and this is the
conventional explanation.

\subsection{X-ray and gamma-ray inferences}

The discovery of X-ray synchrotron emission from SNRs dates from the
early 1980's (Toor 1980; Becker et al.~1980; Reynolds \& Chevalier
1981) but observational and theoretical confusion prevented its wide
acceptance until the unmistakable observations with {\sl ASCA} of the
remnant of SN 1006, demonstrating conclusively that central emission
was thermal while the bright limbs showed lineless spectra well
described by power laws (Koyama et al. 1996).  (See Reynolds 2008 for
a detailed review of high-energy nonthermal emission from SNRs.)  At
this time, four Galactic SNRs show X-ray spectra dominated by
synchrotron emission: G1.9+0.3, SN 1006, G347.3-0.5 (RX J1713.7-3946),
and G266.2-1.2 (``Vela Jr.'' or RX J0852-4622).  Of these, the first
two are young, with symmetric X-ray morphologies, and show thermal
X-ray emission from fainter regions.  SN 1006 is widely accepted to be
the remnant of a Type Ia supernova (evidence includes the presence of
Balmer-dominated optical emission from the forward shock, and the high
Galactic latitude of $15^\circ$); the evidence is not as firm for
G1.9+0.3, but it is also suspected to be a Type Ia remnant
\citep{reynolds08b}.  The other two have much larger angular sizes and
considerably more irregular morphologies \citep{slane99, slane01};
they also show no trace of thermal X-rays, putting severe upper limits
on the presence of any thermal gas \citep{ellison10}.  Both also
contain X-ray point sources, suggesting a core-collapse origin. In
addition to these four objects, all other historical shells (Tycho =
SN 1572, Kepler = SN 1604, Cas A (unseen SN around 1680), G11.2-0.3=SN
386, and RCW 86 = SN 185?) show synchrotron emission in regions,
typically (but not always) ``thin rims'' at the remnant periphery.
All, in addition to SN 1006, also show hard continua in the integrated
X-ray flux seen by the (non-imaging) RXTE satellite \citep{allen99}.
Detections above 8 keV have also been reported by INTEGRAL (Cas A,
Renaud et al.~2006; SN 1006, Kalemci et al.~2006; G347.3-0.5, Krivonos
et al.~2007) and by non-imaging instruments BeppoSAX (Cas A to 80 keV;
Favata et al.~1997) and {\sl Suzaku} HXD (Cas A [to 40 keV], Maeda et
al.~2009; G347.3-0.5 [to 40 keV], Tanaka et al.~2008; and Tycho [to 30
keV], Tamagawa et al.~2009).  Arguments that the radiation is
synchrotron essentially turn on the exclusion of all other possible
processes (see arguments summarized in Reynolds 2008).  Synchrotron
photons at 4 keV energy ($\nu \sim 10^{18}$ Hz) imply electron
energies of order $100 (B/10\ \mu{\rm G})^{1/2}$ TeV.

Power-law fits to the X-ray spectra are much steeper than radio
spectra, indicating that the X-rays come from the cutting-off tail of
the electron distribution.  Three limitations might be imagined for
the maximum energy to which electrons can be accelerated due to DSA:
finite age or size of the remnant, lack of MHD scattering waves above
some $\lambda({\rm max})$, or radiative losses.  The first two
mechanisms would restrict ion acceleration as well, but radiative
losses would affect electrons only.  The cutoffs are likely to be
exponentials or modified exponentials, so that the electron
distribution is roughly given by $N(E) = K E^{-s} e^{-E/E_{\rm max}}$.
In a uniform magnetic field, such a distribution would give rise to a
spectrum $S_\nu \propto \nu^{-\alpha}e^{-\sqrt{\nu/\nu_c}}$
approximately, dropping off rather slowly.  (A more careful
calculation in the case of synchrotron losses predicts a spectrum with
the same exponential factor $e^{-\sqrt{\nu/\nu_c}}$; Zirakashvili \&
Aharonian 2007).  Here $\nu_c \propto E_{\rm max}^2 B$ is the
``rolloff'' frequency corresponding to the peak frequency emitted by
electrons with energy $E_{\rm max}$.

In the cutoff part of the spectrum, we observe the competition between
acceleration and loss rates.  In standard DSA, the time for a particle
to reach a relativistic energy $E$ (from $E_i \ll E$) is given by
$\tau(E) \sim \kappa/u_s^2$ where $\kappa \equiv \lambda_{\rm mfp}c/3$
is the diffusion coefficient.  It is often assumed, but not demanded
by the data, that $\kappa$ is ``Bohm-like,'' that is, that the mean
free path $\lambda_{\rm mfp}$ is proportional to the particle
gyroradius $r_g \equiv E/eB$ (for extreme-relativistic energies).  If
we write $\lambda_{\rm mfp} = \eta r_g$, with $\eta$ the
``gyrofactor,'' depending on the amplitude of scattering turbulence,
then $\eta = 1$ is Bohm diffusion and $\eta = $const.$> 1$ is
``Bohm-like''.  In commonly used ``quasi-linear theory,'' $\eta(E)
\equiv (\delta B/B)^{-2}$ where $\delta B$ is the amplitude of MHD
turbulence resonant with particles of energy $E$.  (So $\eta =$
const. means a particular spectrum of turbulence, equal energy per
unit logarithmic bandwidth).  Bohm diffusion is often taken as a
limiting case, $\delta B \sim B$, but it has been argued that $\delta
B \gg B$ is the possible outcome of some cosmic-ray driven
instabilities (Bell \& Lucek 2001; Bell 2004).  Relevant for using
X-ray spectra as a diagnostic of magnetic fields is that high fields
give short mean free paths, small diffusion coefficients, rapid
acceleration, and higher maximum particle energies.

\begin{figure}
\includegraphics[width=\hsize]{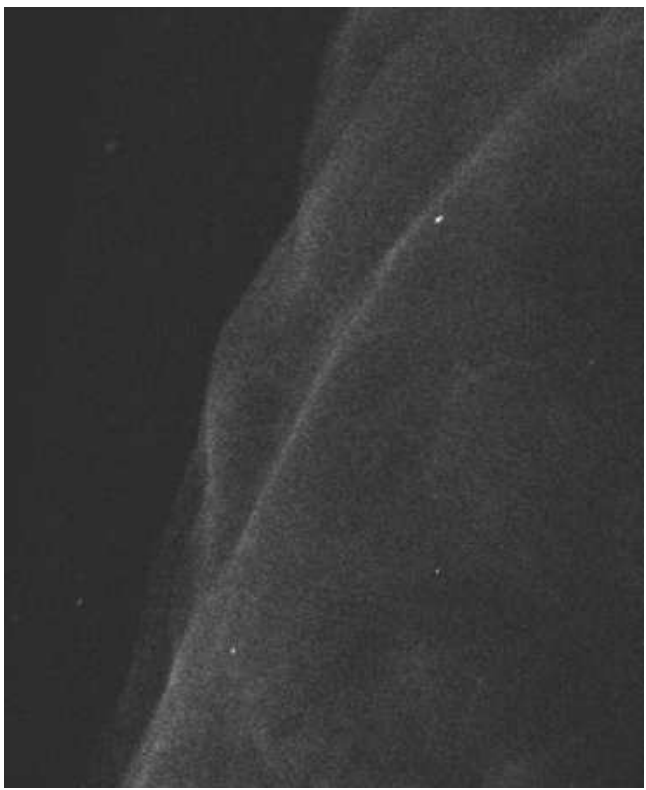}
\caption{Raw {\sl Chandra} counts image of the NE region of the limb
of SN 1006, showing ``thin rims'' of synchrotron radiation (Long et
al.~2003).  The brightness contrast of the very sharp rims is about a
factor of 2. The inner rim has a a clear H$\alpha$ counterpart [Winkler
et al. 2003].}
\label{sn1006fils}
\end{figure}

\begin{figure}
\includegraphics[width=\hsize]{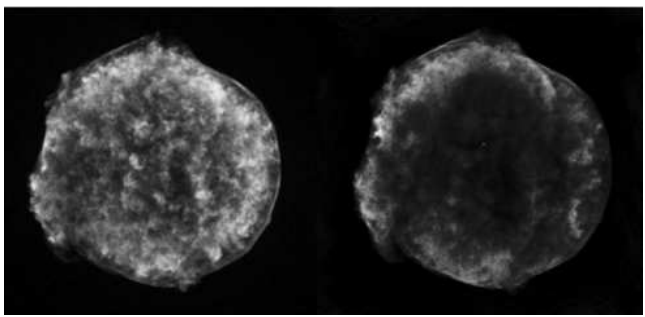}
\caption{Left: {\sl Chandra} image of Tycho's SNR (CXC).  Right: VLA
image at 1420 MHz \citep{reynoso97}.  Note the thin rims in the X-ray
image (e.g., NE quadrant), with some thin radio rims as well.}
\label{tychorx}
\end{figure}

Two classes of argument for high magnetic fields in young SNRs come
from the last decade of observations with high-resolution, high
throughput X-ray observatories: {\sl Chandra}, {\sl XMM-Newton,} and
{\sl Suzaku}.  A morphological argument is based on the commonly seen
phenomenon of ``thin rims,'' in which synchrotron X-ray emission occurs
at remnant peripheries in very narrow tangential features coincident 
with the shock location as inferred, e.g., from H$\alpha$ observations
(see Figure~\ref{sn1006fils}).  \cite{bamba03} and \cite{vink03} argued
that the small radial extent resulted from synchrotron losses on electrons,
to infer strong amplification of $B$:
\begin{equation}
B \sim 200 (u_s/1000\ {\rm km\ s}^{-1})^{2/3} (w/0.01\ {\rm
pc})^{-2/3}\ {\rm G} 
\end{equation}
\citep{parizot06}, where $w$ is the filament width.  Values of $B$
from 60 to 230 $\mu$G have been inferred for Tycho, Kepler, SN 1006,
Cas A, and G347.3-0.5 \citep{parizot06}.  These field strengths apply
only to regions where X-ray synchrotron emission is seen, of course.
An alternative explanation for the disappearance of synchrotron
emission such a short distance behind the shock is that the magnetic
field is primarily turbulent, and decays on a short lengthscale
\citep{pohl05}.  Tests attempting to discriminate between these two
possibilities somewhat favor the radiative-loss explanation, but are
not yet conclusive \citep{cassamchenai07}.  One particular difficulty
is the presence of thin rims in radio images as well (see
Figure~\ref{tychorx}); since radio-emitting electrons have enormously
longer loss timescales, disappearance of the magnetic field is the
only alternative.  It is possible that both processes operate;
theoretical models for strong magnetic-field amplification in modified
shocks produce turbulent magnetic field which might well have some
decay mechanism to compete with electron energy losses.

Another argument for (locally) strong magnetic fields is the discovery
of variations in brightness on timescales of a few years of small
features in X-ray synchrotron emission in Cas A \citep{patnaude07} and
G347.3-0.5 \citep{uchiyama07}.  Again, factors of 2 drop in brightness
on a timescale of years require $B \sim 1$ mG.  Brightness increases
on similar timescales are also observed; demanding acceleration times
of years gives similar estimates for $B$.  However, very strong levels
of magnetic turbulence also naturally predict these kinds of
fluctuations \citep{bykov08, bykov09} without necessarily requiring
quite as high values of $B_{\rm rms}$.  However, the remnant of SN
1006 shows no such small features or brightness variations
\citep{katsuda10}, perhaps as a result of a Type Ia origin with
more-or-less uniform surrounding material.  (The absence of short-term
variations does not imply absence of magnetic-field amplification;
very high $B$ values could be generated in a quasi-steady state behind
the shock, without producing short-term brightness fluctuations.)

\begin{figure}
\includegraphics[width=\hsize]{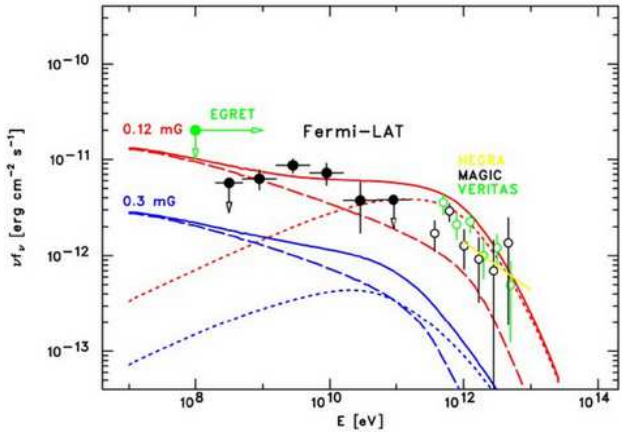}
\caption{Spectrum of GeV emission from Cas A as seen by {\sl Fermi}
\citep{abdo10a}.  The solid curves are leptonic models (dots,
IC; dashes, bremsstrahlung) with magnetic-field values shown.
If the emission is actually hadronic, the magnetic-field strengths
must be larger to suppress the IC component.}
\label{casa}
\end{figure}

New arguments relative to magnetic-field strengths in SNRs have
resulted from the new generation of GeV-TeV observational
capabilities.  The {\sl Fermi} Gamma-Ray Space Telescope has been
mapping the sky between about 0.2 and 300 GeV since its launch in
2008, with gradually improving statistics on any given region of sky.
One important early result is the detection of Cas A between 0.5 and
50 GeV \citep[][see Figure~\ref{casa}]{abdo10a}.  The emission may be
either leptonic, that is, electron bremsstrahlung plus inverse-Compton
upscattering of local photon fields, or hadronic, from the decay into
photons of $\pi^0$ mesons produced in inelastic collisions between
cosmic-ray protons and local thermal gas.  The bremsstrahlung and
hadronic contributions require knowledge of the gas density; the IC
contribution requires knowledge of appropriate photon fields.  The
leptonic model shown in Figure~\ref{casa} assumes a mean density of 26
cm$^{-1}$, which requires a mean magnetic field (averaged over the
entire emitting region of Cas A) of about 0.12 mG.  If the actual
emission is hadronic, the IC contribution must be lower (fewer
relativistic electrons), demanding a larger magnetic field to produce
the observed synchrotron fluxes at radio energies.  (This argument was
originally made based on early upper limits above 100 MeV from COS-B
and SAS-2; Cowsik and Sarkar [1980] derived a lower limit of about 1
mG.)

Similarly, any TeV detection or upper limit places lower limits on the
mean magnetic field in regions containing relativistic electrons.  So
far, four SNRs have been imaged with the HESS air-\v{C}erenkov
telescope array in Namibia.  These include three of the
synchrotron-dominated remnants: SN 1006, G347.3-0.5, and Vela Jr.
(The fourth, G1.9+0.3, is too small for HESS imaging and too faint for
a constraining upper limit). In addition, the (probably) historical
shell SNR RCW 86 has been detected \citep{aharonian09}.  In all cases,
the TeV morphology tracks the X-ray morphology surprisingly closely.
However, detailed attempts to model the emission as either due to
hadronic or to leptonic processes run into significant problems.  For
hadronic models, required target gas densities seem to exceed
measurements or upper limits on local thermal gas, while simple
leptonic models do not do a good job reproducing the spectral shape
and may also require implausibly small filling factors of magnetic
field.  However, lower limits to $\langle B\rangle$ do not depend much
on the detailed modeling.  In SN 1006 \citep{acero10}, $B \gapprox 30\
\mu$G is required; for G347.3-0.5, a leptonic model \citep{lazendic04}
demands a filling factor of magnetic field of only 1\% to avoid
overpredicting synchrotron emission.  Within that 1\% of the volume,
the field is a modest 15 $\mu$G.  (These estimates are relatively
insensitive to model details; even a model that does not use DSA at
all as a particle acceleration mechanism, but rather stochastic
acceleration in the downstream region, fits the most recent GeV and
TeV data for G347.3-0.5 with a leptonic model with $B \sim 12\ \mu$G;
Fan, Liu, \& Fryer 2010).  Hadronic models typically invoke much
higher fields (Zirakashvili \& Aharonian [2010] find $B = 127\ \mu$G
in G347.3-0.5).

HESS has also detected several other sources coincident with shell
SNRs, such as CTB37B (HESS J1713-381; Aharonian et al.~2008) and HESS
J1731-347 (Acero et al.~2009).  These may be further members of the
nonthermal-X-ray class, older SNRs interacting with molecular clouds,
or PWNe.  They await further observational clarification.

The question of magnetic-field amplification is intimately connected
with that of efficient particle acceleration through the proposals of
Bell \& Lucek (2001) and Bell (2004) that cosmic-ray driven
instabilities can greatly increase the magnetic-field strength.
Evidence for high magnetic fields is then taken as indirect evidence
for an energetically significant component of cosmic rays (which must
of necessity be ions) accelerated in the shock.  This argument can be
reversed: if evidence is found for efficient ion acceleration, then
magnetic fields are likely to be amplified.

\subsection{Obliquity dependence}

The nature of magnetic-field amplification at strong shocks is a
subject of intense theoretical activity.  Observations can provide
powerful constraints on some aspects of such amplification.  In
particular, two SNRs, SN 1006 and G1.9+0.3, show very simple
bilaterally symmetric morphology which may contain clues to the
dependence of particle acceleration and magnetic-field amplification
processes on the shock obliquity angle $\theta_{\rm Bn}$ between the
mean upstream magnetic field direction and the shock velocity.
Section~\ref{bocch} discusses in detail a model for the radio morphology
of SN 1006, as representative of this bilaterally symmetric class of
SNRs (BSNRs).

\begin{figure}
\includegraphics[width=\hsize]{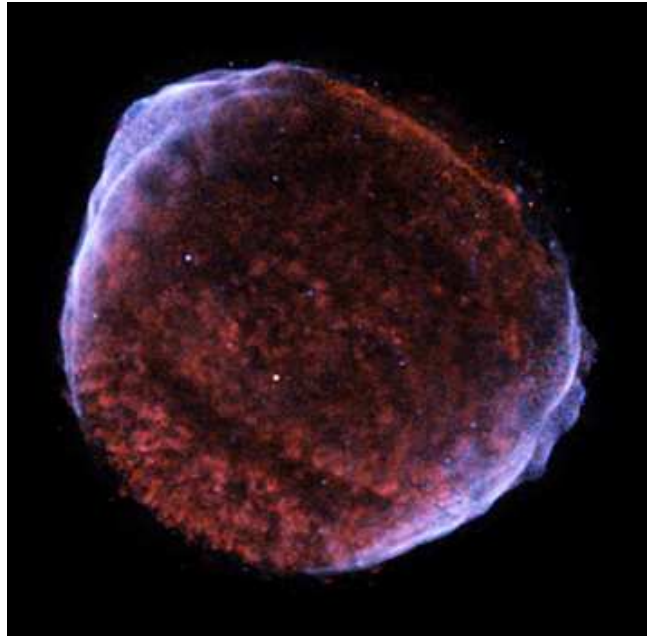}
\caption{SN 1006 with {\sl Chandra} (NASA/CXC).  Red, 0.50 -- 0.91 keV;
cyan, 0.91 -- 1.34 keV; blue, 1.34 -- 3.00 keV.  Blue-white rims are
line-free synchrotron emission; red interior is dominated by oxygen
ejecta.}
\label{sn1006x}
\end{figure}

\begin{figure}
\includegraphics[width=\hsize]{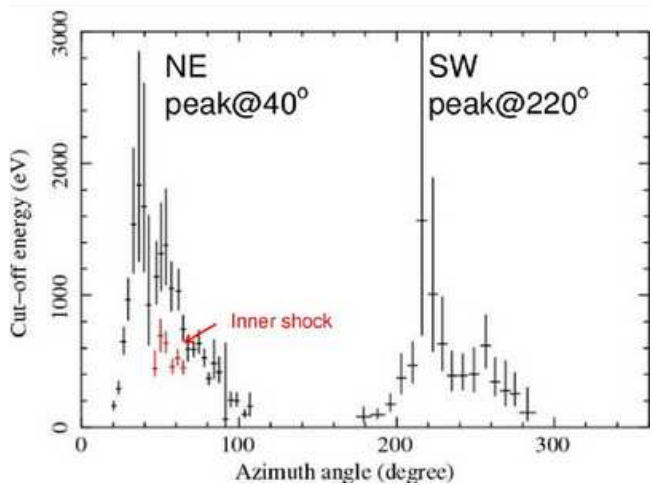}
\caption{Rolloff frequencies measured in $10''$ deep (radially; much
larger in the azimuthal direction) rectangular regions around the
outer edge of SN 1006 (red points: an interior rim) (Katsuda et
al., in preparation).  The exposure time is much longer in the NE region, so
boxes are smaller.}
\label{sn1006azim}
\end{figure}

\begin{figure}
\includegraphics[width=\hsize]{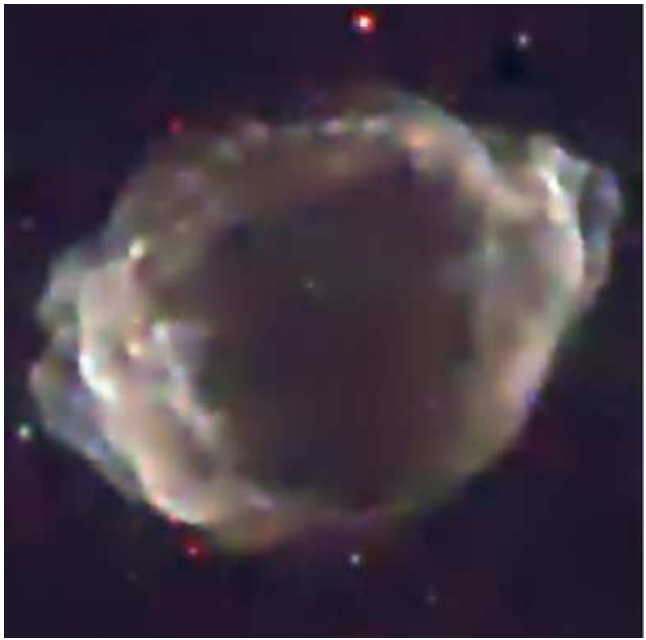}
\caption{G1.9+0.3, the Galaxy's youngest SNR ({\sl Chandra}; Borkowski
et al.~2010).  Red,  1 -- 3 keV; green, 3 -- 5 keV; blue, 5 -- 8 keV.
Except for faint thermal emission in the center and north, the 
emission is dominantly synchrotron.}
\label{g1.9}
\end{figure}

\begin{figure}
\includegraphics[width=\hsize]{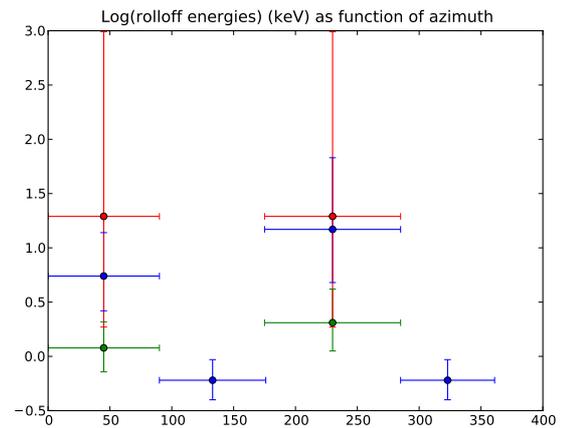}
\caption{Variations of $\nu_{\rm roll}$ with radius and azimuth, for
G1.9+0.3.  Red: ``Ears'' extending beyond the main shell to E and W.
Blue: main shell.  Green: Just interior to bright regions of main
shell.  Note the similar trend at much poorer resolution to that seen
in Figure~\ref{sn1006azim}.}
\label{g1.9azim}
\end{figure}

The most natural explanation for the bilateral symmetry evident in
Figures~\ref{sn1006x} and \ref{g1.9} is the variation in $\theta_{\rm
Bn}$ one would encounter at different locations as a spherical shock
encounters a roughly uniform magnetic field.  If the field lies in the
plane of the sky, the range in obliquities would be 0 to $\pi/2$; if
there is a line-of-sight component, the range drops until it is zero
for {\bf B} along the line of sight.  Later in this chapter we
describe a detailed calculation of the radio synchrotron morphologies
resulting from different assumptions about the nature of the upstream
magnetic field and of particle acceleration. Here we focus on the X-ray
spectral variations each remnant displays.  


A simple model adequately characterizing most synchrotron X-ray emission
from shell SNRs is the ``srcut'' model in the XSPEC software package,
which is the spectrum emitted by a power-law electron distribution
with an exponential cutoff.  The frequency corresponding to the
cutoff energy $E_{\rm max}$ is $\nu_{\rm roll}$, the ``rolloff'' frequency
(so called because the synchrotron spectrum from this distribution actually
steepens rather gradually).  $\nu_{\rm roll}$ serves as a good indicator
of the maximum energy reached by electrons at different locations, although
depending on the mechanism limiting that maximum energy, $\nu_{\rm roll}$
has different dependencies on physical parameters.

If the acceleration rate of particles varies with shock obliquity
(independently of any obliquity-variations of the magnetic field), we
can parameterize such effects in terms of a quantity $R_J$:
$R_J(\theta_{\rm Bn}, \eta, r) \equiv \tau(\theta_{\rm
Bn})/\tau(\theta_{\rm Bn} = 0)$.  (This means that the value of $\eta$
is that for $\theta_{\rm Bn} = 0$; any obliquity-dependence is
subsumed into $R_J$.)  We scale to typical values for young SNRs:
$u_{8.5} \equiv u_{\rm sh}/3000$ km s$^{-1}$; $t_3 \equiv t/1000$ yr;
$B_{10} \equiv B/10 \ \mu{\rm G}$; and $\lambda_{17} \equiv
\lambda_{\rm max}/10^{17}$ cm.  Then the values of $\nu_{\rm roll}$
for each limiting mechanism obey \citep{reynolds08a}
\begin{eqnarray}
h\nu_{\rm roll} ({\rm age}) &\sim& 0.4\, u_{8.5}^4\, t_3^2\, B_{10}^3\, (\eta R_J)^{-2}\ {\rm keV} \\
h\nu_{\rm roll} ({\rm esc}) &\sim& 2 \,B_{10}^3\, \lambda_{17}^2 \ {\rm keV} \\
h\nu_{\rm roll} ({\rm loss}) &\sim& 2\, u_{8.5}^2\, (\eta R_J)^{-1}\ {\rm keV}.
\label{roll}
\end{eqnarray}
The effective limiting mechanism is the one producing the lowest value
of $E_{\rm max}$; this may vary around the remnant periphery.
Note that if radiative losses limit electron acceleration, 
the resulting $\nu_{\rm roll}$ is independent of the magnetic-field
strength. 

\begin{figure}
\includegraphics[width=\hsize]{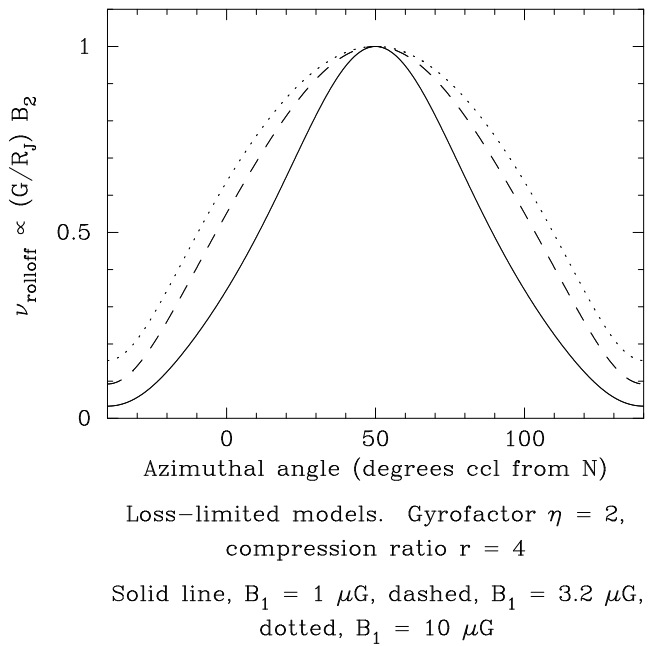}
\caption{Variation of $\nu_{\rm roll}$ with azimuth
\citep{reynolds98}, according to the anisotropic diffusion model of
Jokipii (1987).  The required amount of variation of $\nu_{\rm roll}$
can be produced, but only if the shock is perpendicular at the bright
limbs. $B_1$ and $B_2$ are the upstream and downstream magnetic-field
strengths, and $G$ is the ratio of an electron's energy-loss time in
a field $B_1$ to that in the actual obliquity-dependent combination
of $B_1$ and $B_2$ as it scatters back and forth across the shock;
see Reynolds 1998.}
\label{rj}
\end{figure}

Figure~\ref{sn1006azim} shows the variation in $\nu_{\rm roll}$ around
the periphery of SN 1006 (Katsuda et al., in preparation); similar
data for G1.9+0.3 (Fig.~11) at much lower resolution
\citep{reynolds09} are shown in Figure~\ref{g1.9azim}.  Variations by
large factors occur.  Equations 3 -- 5 above show that achieving
variation by over an order of magnitude is not easy, especially for
loss-limited acceleration.  Even in the absence of any
cosmic-ray-driven magnetic-field amplification, one would expect a
variation in $B$ by about the shock compression ratio (4 or more)
simply due to compression of the tangential component of $B$ due to
flux-freezing.  While there are no comprehensive predictions,
calculations of magnetic-field amplification normally assume parallel
shocks ($\theta_{\rm Bn} \sim 0$), with the process being less
effective at higher obliquities.  The strong $B$-dependence for age or
escape-limited acceleration could easily explain the observations --
but with the serious side effect of predicting that the resulting
maximum energies ($E_{\rm max} \sim$ 10 -- 100 TeV) would apply to
ions as well as electrons, putting at least these two SNRs out of the
running for producing cosmic rays up to the ``knee'', the slight
steepening in the integrated spectrum of Galactic cosmic rays at
Earth.  Loss-limited maximum energies do depend on the shock speed,
but the required variation by over 30 in $\nu_{\rm roll}$ would demand
an implausible factor of 5 variation in the current shock velocity.
Anisotropic diffusion \citep{jokipii87} could produce more rapid
acceleration at quasi-perpendicular shocks than quasi-parallel ($R_J
(\theta_{\rm Bn}) < R_J(0)$), easily explaining the required amount of
variation (Figure~\ref{rj}).  However, most workers believe that this
effect may not occur in highly turbulent media. In addition, electron
injection into the DSA process may be more difficult at
quasi-perpendicular shocks.  As we show later in this chapter, the
radio morphology may be more accurately reproduced with a
quasi-parallel model, leaving the explanations of
Figures~\ref{sn1006azim} and~\ref{g1.9azim} as an open question.

\subsection{Summary}

We summarize here the inferences on magnetic-field strength and geometry
described above.

\begin{enumerate}

\item From radio observations, equipartition values of magnetic field
strength are in the $\sim 10\ \mu$G range, but there is little
physical motivation to assume equipartition.

\item Radio polarization studies show that in young SNRs, the magnetic
field is largely disordered, with a small radial preponderance.  In
older, larger SNRs, the field is often disordered but sometimes
tangential.

\item Curvature (spectral hardening to higher frequency) is observed
in the radio spectra of Tycho and Kepler.  A nonlinear shock
acceleration model can explain this with magnetic field strengths of
0.1 -- 1 mG (average over the emitting regions).

\item Thin rims of X-ray synchrotron emission in a few young remnants
require $B \sim 50 - 200\ \mu$G in the rims, if they are due to
synchrotron losses on down-stream-convecting electrons.  However, thin
radio rims are sometimes seen as well; they require that the magnetic
field disappear somehow, presumably because it is a wave field which
damps.

\item Brightening and fading of small X-ray synchrotron features in
G347.3-0.5 and Cas A require $B \sim 1$ mG, if they represent
acceleration and loss times for electrons.  Fields smaller by a factor
of several are possible if the fluctuations are due to strong magnetic
turbulence.

\item Large azimuthal variations in the rolloff frequency in SN 1006
and G1.9+0.3 are difficult to explain for a conventional picture of
loss-limited acceleration in parallel shocks.

\item For Cas A, the detection at GeV energies with {\sl Fermi}
requires $B \gapprox 0.1$ mG to avoid overproducing the GeV emission
with electron bremsstrahlung.

\item TeV emission seen in four shell SNRs is not well explained by
either leptonic or hadronic processes.  However, if it is hadronic,
the magnetic fields implied are of order 100 $\mu$G, while leptonic
models require much lower fields.

\end{enumerate}



\section{Magnetic fields in pulsar-wind nebulae}

\subsection{Introduction}

All young pulsars are slowing down in their rotation. The
corresponding rate of change of rotational kinetic energy is enormous,
typically in the range $\dot{E} = 10^{32}-10^{39}$ ergs s$^{-1}$.  In
most cases only a negligible amount of energy is contained in the
electromagnetic radiation corresponding to the neutron star's observed
pulsations.  Rather, the bulk of the rotational power is usually
deposited into a relativistic magnetized particle wind that flows
outward from the pulsar.

\begin{figure}
\includegraphics[width=\hsize]{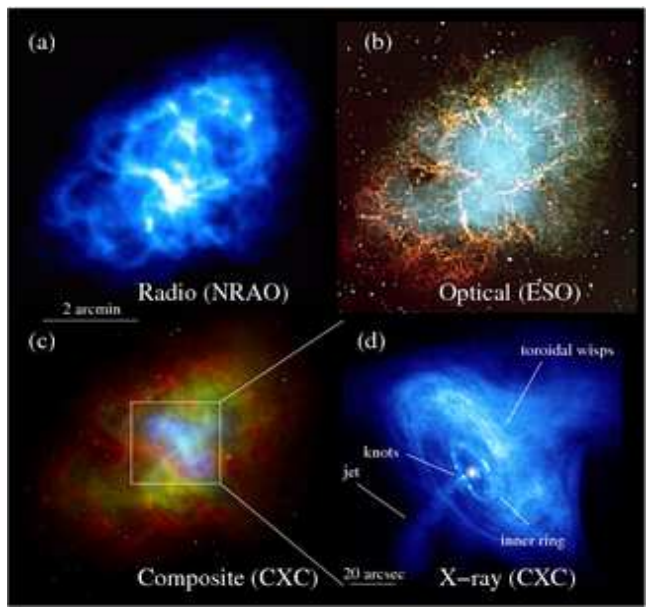}
\caption{Images of the Crab Nebula.}
\label{crab}
\end{figure}

External pressure will cause this wind to abruptly decelerate at a
termination shock. Beyond the termination shock, the pulsar wind
thermalizes in pitch angle and radiates synchrotron emission,
resulting in a pulsar wind nebula (PWN). The best known PWN is the
Crab Nebula (Fig.~\ref{crab}), powered by the central young pulsar
B0531+21.  The Crab Nebula radiates synchrotron emission across the
electromagnetic spectrum. In X-rays, its extent is small, reflecting
the relatively short synchrotron lifetimes of electrons emitting at
these energies.  The optical synchrotron emission is larger,
reflecting the longer lifetimes of the corresponding electron
population. And finally, the radio emission shows the full extent of
the source, with a radiative lifetime exceeding the age of the source
($\sim1000$~years).

\subsection{Observational Properties}

PWNe and SNRs have a variety of common properties: they both result
from a supernova explosion, and they both are characterized by their
polarized radio synchrotron emission.  This has led to considerable
confusion in the literature.  For example, the Crab Nebula is often
described as a ``supernova remnant'', despite the shell or blast-wave
that would normally correspond to a SNR remaining as yet unidentified
in this source \citep{fkcg95,sgs06,tsls09}.  Some PWNe have for
historical reasons been cataloged as ``filled-center supernova
remnants'' \citep{green09}, while others, discovered more recently,
have been excluded from such lists.  As we describe below, a young
SNR may contain a PWN, a combination often called a composite
remnant (e.g., G21.5-0.9; Fig.~\ref{g21}). 

The key differences between SNRs and PWNe are as follows:
\begin{itemize}
\item {\bf Energy Source:} SNRs result from an essentially instantaneous
deposition of energy, in the form of a blast wave driven into the
ISM by a supernova explosion. In contrast, PWNe have a continuous
power source, the bulk relativistic flow of electron/positron pairs
from an energetic neutron star.

\item {\bf Radio Morphology:} As a direct result of their
differing sources of energy, SNRs and PWNe have distinct
radio morphologies. SNRs are usually limb-brightened shells
of synchrotron emission, while PWNe are typically amorphous
or filled-center synchrotron nebulae, brightest at the pulsar's
position. 

\item {\bf Radio Spectral Index:} SNRs usually have relatively steep
radio spectral indices, $\alpha \approx 0.3-0.8$, as shown in
Figure~\ref{spix}. In contrast, PWNe have spectral indices in the
range, $\alpha \approx 0-0.3$, which is too flat to be explained
by simple models of diffusive shock acceleration
\citep[e.g.,][]{ato99,fb07,tt10}.

\item {\bf Angular Extent:} SNRs are long-lived objects with a wide
range of sizes, with angular extents ranging from $\sim 1'$ to
$>5^\circ$. PWNe are usually relatively small, with sizes
in the range $10''$ to $30'$, although a few older PWNe may 
be significantly larger (e.g., Vela X; Fig.~\ref{vela}).  

\item {\bf Fractional Polarization:} At radio frequencies
near 1~GHz, SNRs typically have modest amounts
of linear polarization, at the level of 5\%--10\%. 
PWNe usually have very well-organized magnetic
fields, with correspondingly higher polarization
fractions, in the range 30\%--50\%.
\end{itemize}


\subsection{PWN Evolution}

Theoretical expectations predict three main phases in the evolution of
a PWN \citep{rc84,che98,che05}, governed in turn by:
\begin{enumerate}
\item the expansion of the PWN;
\item interaction of the PWN with the surrounding SNR;
\item the motion of the pulsar powering the PWN.
\end{enumerate}

Each of these phases is discussed in more detail below.

\subsubsection{Phase 1: Expansion into Unshocked Ejecta}

\begin{figure}
\includegraphics[width=\hsize]{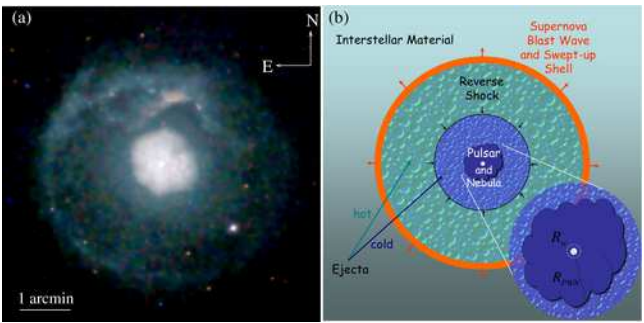}
\caption{Left: Deep {\it Chandra} image of the composite SNR G21.5-0.9 
\citep{matheson05}. The PWN is the bright central condensation.
Right: Schematic structure of a composite SNR containing a PWN.}
\label{g21}
\end{figure}

At early stages in a PWN's evolution, the pulsar's spin-down
luminosity remains relatively constant, so that there is a steady
injection of energy into the PWN. As a consequence, the PWN expands
supersonically into the surrounding low-pressure environment, with the
radius, $R$, of the PWN evolving with time, $t$, as $R \propto
t^{6/5}$ \citep{che77}.  Thus the PWN drives a shock into the inner
edge of the expanding SN ejecta (see Fig.~\ref{g21}).

Many pulsars are born with high space velocities, up to and beyond
1000~km~s$^{-1}$. However, because the sound speed inside the PWN is
much higher than this, the PWN stays relatively centered on the
pulsar, and has a quasi-spherical appearance.

Inside the PWN, there are two broad zones. Close to the pulsar is the
unshocked wind zone, in which particles and Poynting flux flow freely
outward from the central star. The pressure in this wind drops with
increasing distance from the pulsar, until we reach a termination
shock at which the wind is confined by external pressure.  At this
shock the wind particles are accelerated up to even higher Lorentz
factors, and have their pitch angles scattered so that they can emit
synchrotron radiation.  Observations of this emission allow us to
infer the composition of the wind at and immediately downstream of the
shock. Specifically, the parameter $\sigma$ corresponds to the ratio
of electromagnetic energy to particle energy at the wind shock, and is
usually much less than one \citep{kc84a,gak+02}: the shocked wind is
particle-dominated ($\sigma \ll 1$), in contrast to the unshocked wind
which is expected to be radiation-dominated ($\sigma \gg 1$).

Beyond the termination shock is the second main zone, the emitting
region of the PWN, in which the bulk flow continues to decelerate to
match the external boundary conditions, and where synchrotron emission
is produced.

The morphology of a PWN in this early phase of evolution has been
spectacularly revealed by X-ray images of the Crab Nebula taken by the
{\em Chandra X-ray Observatory}\ \citep{wht+00,hmb+02}. These
observations show that the Crab Nebula is dominated by a bright X-ray
torus. The inner part of this torus is bounded by a bright X-ray ring,
thought to correspond to the wind termination shock.  To the south of
the pulsar, an X-ray jet runs along the axis of the torus, apparently
originating very close to the central pulsar before curving slightly
at large distances from the center. A faint counter-jet can also be
seen to the north.


The torus is thought to result from the Poynting flux concentrated in
equatorial regions of the system by the wound-up magnetic field.
However, a key point, first noted by \cite{lyu02b}, is that the
termination shock radius should decrease with increasing angle from
the equator. The axial jet seen for the Crab Nebula is therefore not a
jet originating from the pulsar itself, but is part of the post-shock
wind \citep[e.g.,][]{bk02}.

Relativistic magnetohydrodynamic simulations show that at
mid-latitudes, the external pressure can reverse the post-shock
particle flow. The wind flows back along the surface of the
termination shock toward the poles \citep{kl03,dab04,dvab06}, at which
points it is collimated into jets by hoop stress \cite[the
``toothpaste effect;][]{ckbh09}. Simulations thus can do a reasonable
job of reproducing the morphological properties of the Crab Nebula, as
a result of a latitude-dependent termination-shock radius with a flow
reversal.

Recently published optical polarimetry of the Crab Nebula using the
{\em Hubble Space Telescope}\ has revealed spectacular details of this
PWN's large-scale magnetic field \citep{hes08}. A great deal of
fibrous structure seen in polarization corresponds to the local field
direction, while the torus of the nebula appears to be confined by a
poloidal component of the field.


\subsubsection{Phase 2: Interaction with Reverse Shock}

A PWN moves into a new phase of evolution when its outer boundary
collides with the reverse shock from the SNR in which it is embedded.
This typically occurs after $\sim7000$~years for the typical case of a
$10^{51}$~erg supernova ejecting 10 solar masses into an ambient ISM
of density 1 atom cm$^{-3}$ \citep{rc84}. The collision with the
reverse shock initially compresses the PWN, and the system may
reverberate for a couple of cycles before establishing a new
equilibrium \citep{bcf01,vagt01,gsz09}

\begin{figure}
\includegraphics[width=\hsize]{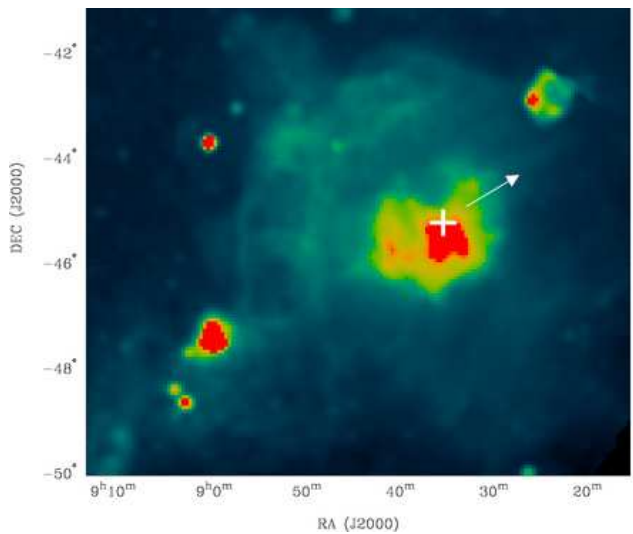}
\caption{2.4 GHz Parkes image of the Vela supernova remnant
\citep{duncan96}.  The white cross indicates the pulsar, and the arrow
its proper motion \citep{dodson03}; the fact that the pulsar is
neither at nor moving away from the PWN's center indicates that the
reverse-shock interaction has take place.}
\label{vela}
\end{figure}

The compression of the PWN raises the internal magnetic field,
resulting in substantial synchrotron burn-off at high energies.  The
X-ray extent of a PWN in this phase can therefore be rather small.
Furthermore, if the pulsar has a high space velocity or if the
surrounding ISM is inhomogeneous, then the reverse shock does not
collide with all parts of the PWN at the same time.  As a result of
this asymmetric collision, the pulsar can end up substantially offset
from the center of its PWN, and the PWN's radio morphology can take on
a chaotic or filamentary appearance \citep{bcf01,vdk04} (see
Fig.~\ref{vela}).

\subsubsection{Phase 3: Supersonic Motion}

\begin{figure}
\includegraphics[width=\hsize]{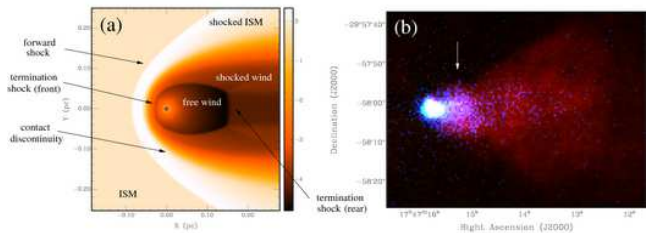}
\caption{Left: Hydrodynamic simulation of a pulsar bow shock.  The
pulsar (asterisk) is moving from right to left with Mach number 60.
Right: {\sl Chandra} image (blue) and VLA image (red) of G359.23-0.82
(the ``Mouse''), the bow shock associated with PSR J1747-2958
\citep{gvc+04}.  The white arrow indicates a structure which
may be the termination shock.}
\label{mouse}
\end{figure}

Since a pulsar moves ballistically while the surrounding SNR
decelerates in its expansion, any pulsar with a significant space
velocity will eventually begin to move toward the SNR's rim.  For a
SNR in the Sedov phase, the sound-speed in the remnant's interior
drops with increasing distance from the center. Specifically,
\cite{vag98} has shown that for a Sedov SNR, the pulsar's motion will
become supersonic when it has moved 68\% of the way to the SNR's rim,
which occurs at 50\% of the total time taken for the pulsar to escape
the SNR completely. Once the pulsar becomes supersonic, a new,
bow-shock, PWN is formed, which no longer expands and which is
externally confined by the ram pressure resulting from the pulsar's
motion.  Eventually the pulsar escapes the SNR completely, and now
drives a bow shock through the ambient ISM
\citep{buc02a,cc02,gvc+04,bad05} (see Fig.~\ref{mouse}).


\subsubsection {Measuring PWN magnetic fields}

We can infer the strength of the magnetic field inside PWNe through
various techniques: by assuming equipartition between particles and
magnetic fields averaged over an assumed synchrotron emitting volume
\citep[e.g.,][]{hss+95}, by measuring the radius of the termination
shock from a high-resolution X-ray image and combining this with a
physical model for the flow to estimate the field value at this
location \citep[e.g.,][]{gak+02}, or through observations of inverse
Compton emission in GeV or TeV gamma-rays (e.g., MSH 15-5{\it 2}: Abdo
et al.~2010b, Aharonian et al.~2005).  De Jager \& Djannati-Ata\"i
(2008) review GeV/TeV emission from PWNe.
These approaches show that PWNe can have a wide range of nebular
magnetic fields, from $\sim$5~$\mu$G to $>$1~mG. The magnetic field
strength generally depends strongly on $\dot{E}$, $t$ and $\sigma$, as
well as on the pulsar braking index.

A variety of magnetic field orientations have been inferred in PWNe
from the angles of their radio polarization. Some PWNe show a broadly
toroidal magnetic field, some show a radial magnetic field, and in
others there is a complex or tangled appearance. \cite{kru06} have
suggested that all PWNe have predominantly toroidal magnetic fields,
but that these different polarization patterns result from differences
in viewing angle. Toroidal polarization vectors, such as seen for the
PWN G106.6+2.9, result when the observer looks along the pulsar's spin
axis \citep{kru06}. Radial fields, such as seen in the PWN G21.5--0.9
\citep{rei02b} correspond to a viewing angle perpendicular to the spin
axis.  And more complex magnetic field morphologies, as for the PWN
3C~58 \citep{ww76}, occur for oblique viewing angles.  The complex
polarization properties of systems like 3C~58 may also occur as a
result of kink instabilities, which can produce sheared loops of
magnetic field throughout the nebula \citep{beg98,shvm04}.

There are only limited observations of magnetic fields in pulsar bow
shocks. The theoretical expectation is that the system forms a
toroidal field near the apex of the bow shock, but that the magnetic
field is collimated in the trailing magnetotail. Radio polarimetry
matching this prediction has recently been observed by \cite{ngcj10}
for the bow-shock PWN powered by PSR~J1509--5850 . However, other
magnetic field geometries have also been seen \citep[e.g.,][]{yg05},
suggesting that the magnetic field structure of bow-shock PWNe may
depend on whether the pulsar's spin axis aligns or mis-aligns with its
velocity vector \citep[e.g.,][]{vmc+07}.

\subsection{Outstanding Issues and Future Work}

There are many aspects of PWNe which are still active areas of
investigation. Two particular areas of focus are as follows.

First, several PWNe have now been observed to show substantial time
variability in their morphology and intensity
\citep[e.g.,][]{pksg01,hmb+02,dgap06,dmcb07}. These observations have
provided a rich suite of information, but we now need relativistic
magnetohydrodynamic simulations to fully understand the flow
conditions and instabilities that produce these effects
\citep{kl04,dab04,bad05, buc10}.

Second, while we have a multitude of observations of pulsar winds
after they have been shocked, we have very limited data on the
composition and transportation mechanism in the unshocked wind zone.
Targeted observations and modeling of unique systems such as
PSR~J0737--3039A/B (the ``double pulsar'') may provide vital insights
in this regard \citep{mll+04,lt05}.

\section{Constraints on the local interstellar magnetic field from
radio emission of SN 1006}
\label{bocch}

\subsection{Introduction}

The radio morphology of supernova remnants (SNRs) may be very informative
of the conditions of the magnetized environments in which the blast-wave
expansion occurs and, in particular, on the acceleration processes
at the shock front which give rise to the energetic
electrons ultimately responsible of the synchrotron emission in the
radio and (possibly) X-ray band. In this context, the 
bilaterally symmetric or barrel-shaped BSNRs
\cite{kc87}, \cite{gae98}) are considered ideal laboratories, because
their morphology is definitely the result of the lack of small scale
inhomogeneties which may confuse the interpretations.  A point-like
explosion in a uniform magnetized medium with constant strength and
direction of {\bf B} should give rise to a symmetric BSNR whose bright
limbs are located where the magnetic field is parallel or
perpendicular to the shock speed, if the injection efficiency is
greatest where the shock is quasi-parallel or
quasi-perpendicular/isotropic, respectively, and if {\bf B} lies in
the plane of the sky, whereas different configurations occur at
different aspect angles (\cite{fr90}).

However, in real life, BSNRs are often asymmetric, and therefore
\citet{obr07} (hereafter Paper I) have recently generalized the study
of \citet{fr90} to the cases in which the explosions occur in a large
scale gradient of density or magnetic field, showing that this model
is able to reproduce most of the asymmetries observed in real
BSNRs. In particular, the radio morphology loses one axis of symmetry,
and the limbs are not equally bright (if the gradient runs across the
limbs) or they are not opposite and they converge on the side in which
the density or the magnetic field is increasing (if the gradient runs
between and parallel to the limbs).

It is clear that morphology of BSNRs is tightly coupled to the
magnetized environments in which the shock expands, and it is of
particular interest to note here the preference of BSNR symmetry axes
to be oriented parallel to the Galactic plane, as reported by
\citet{gae98}. It seems therefore possible to study their morphology
to derive the geometry of the surrounding magnetic field, thus
shedding more light on the microphysics of the particle acceleration
processes at the shock front.

The remnant of SN 1006 seems to be the object in which this kind of
study may be most fruitful. The uniform environment and the bright
limbs visible in most of the electromagnetic spectrum make it a real
case study in the field of particle acceleration mechanisms in strong
shocks.  Indeed, \citet{rbd04}, using a simple and powerful
geometrical argument applied to the \xmm\ X-ray image of SN 1006, based
on the ratio between the central and the rim luminosity, showed that
if the remnant is cylindrically
symmetric, the bright limbs are likely to be polar caps (instead of an
equatorial belt) and that, therefore, the magnetic field is oriented
perpendicular to the bright limbs, in the NE-SW direction. This
argument seems to break the dichotomy between the two competing
scenarios of the dependence of the electron injection efficiency 
or electron acceleration rate
on $\theta_{\rm Bn}$,
preferring a situation in which the injection is most efficient or
acceleration most rapid when the field is along the shock speed
(quasi-parallel scenario) over the situation in which the field is
perpendicular (quasi-perpendicular scenario, see \cite{fr90})\footnote{Since X-ray synchrotron brightness depends on both the
efficiency of electron injection into the acceleration process, and on
the rapidity of acceleration to high energies, studies of the X-ray
morphology involve a combination of both possible effects, while radio
studies are insensitive to acceleration-rate issues, because
acceleration to the GeV energies required for radio emission is always
extremely rapid compared to evolutionary timescales.  Thus variations
of radio morphology with obliquity point to electron injection physics
alone.}.

On one hand, these findings were somehow expected since \citet{vbk03}
already pointed out that in SN 1006 the injection should be maximum at
parallel shocks. However, in the light of the uncertainties related to
the details of the acceleration processes, several authors still
considered the quasi-perpendicular scenario a viable option:
\citet{fr90} argued against the quasi-parallel scenario pointing out
that quasi-parallel models often give rise to unobserved morphologies
in the radio band; \citet{yyt04} still considered both models to
explain the observed width of selected filaments of SN 1006 observed by
\citet{bamba03}, whereas \citet{ah07} developed a quasi-perpendicular
model which agrees very well with the same data. Moreover, the same
simple geometrical argument of \citet{rbd04}, if applied to the radio
image, would be in agreement with the equatorial belt
(cfr. $R_{\pi/3}=0.7$ in Sect. 3.2 of \cite{rbd04}). This discrepancy
has never been explained in the literature and remain one of the most
intriguing open issues in the comparison between models and
observations.

Recently, \citet{pdc09} have further investigated this issue, showing
that, in the framework of a simple model of SN 1006 in terms of a
point-like explosion occurring in a uniform density and uniform
magnetic field medium, there is no way to reconcile the quasi-parallel
scenario and the SN 1006 morphology as observed in the radio band. This
contradiction between the radio morphology (suggesting
quasi-perpendicular injection efficiency scenario) and X-ray
morphology (suggesting quasi-parallel scenario) has prompted us to
investigate the effects of non-uniformity of the environment on the
observed properties of this remnant, capitalizing on the work of
\citet{obr07} on asymmetric BSNRs.


In Sect. 3.2, we will briefly describe the MHD model we have used to
reproduce the remnant of SN 1006, which include a small gradient of the
magnetic field. In Sect. 3.3, we introduce the methodology for the
comparison between modeled and observed radio images of SN 1006, while
in Sect. 3.4 we will discuss the results of the comparison, showing
that, not only can the model invoking a gradient of $|{\bf B}|$ 
reconcile, for the first time, the radio and X-ray morphology of the
remnant, but it also provides stringent constraints on the overall
geometry of the field in the vicinity of the remnant.

\subsection{The model}

\begin{table}
\caption{MHD models of SN 1006 used in this work}
\label{models}
\medskip
\centering\begin{minipage}{8.7cm}
\begin{tabular}{lcccc} \hline
  Name & $\nabla |{\bf B}|$\footnote{The 
  relative variation of the modulus of the magnetic field over a 
   scale of 10 pc} \\ \hline

UNIFORM\_B &  0 \\
GRAD1      &  1.4 \\
GRAD2      &  1.5 \\
GRAD3      &  2.0 \\
GRAD4      &  4.7 \\
\hline
\end{tabular}
\end{minipage}
\end{table}

Since there is accumulated evidence that the density around SN 1006 is
fairly constant, with the exception of the NW sector where an
encounter with a dense cloud is occurring (\cite{abd07},
\cite{mbi09}), we argue that a model with a $|{\bf B}|$ gradient is
more appropriate to describe the asymmetries of the limbs of SN
1006. The remnant was modeled as a point-like explosion of 1.4 solar
masses of ejecta having a kinetic energy of $E=1.3\times 10^{51}$ erg
occurring in a uniform density medium with $n_0 = 0.05$ cm$^{-3}$. The
ISMF has a value of 30 $\mu$G in the environment of the explosion site
and it is assumed to have a gradient along the Y axis.  {\bf Such a
high value (for the location of SN1006 in the Galaxy) has been chosen
in order to take into account as a first approximation the effects of
magnetic-field amplification, to have a post-shock $|{\bf B}|$ of
the order of $\sim 100 \mu$G in our simulation, in agreement with
observations of some SN1006 filaments. Moreover, the magnetic field
has} a gradient along the Y axis.  The direction of the magnetic field
is in the XY plane.  $\nabla |{\bf B}|$ values used in our simulations
are reported in Table \ref{models}. The gradient in the initial
conditions has been modeled with a dipole located along the Y axis,
exactly as in \citet{obr07}, to which the reader is referred for
further details on the model and numerical code used for the
simulations. The simulations were stopped at $t=1002$ yr, checking
that the shock velocity and remnant radius are compatible with the
observed values (4600 km s$^{-1}$ and 9.6 pc, respectively).

\subsection{Comparison between models and observations}

\subsubsection{The angles defining the viewing geometry}

\begin{figure}
  \centerline{\includegraphics[width=9.0cm]{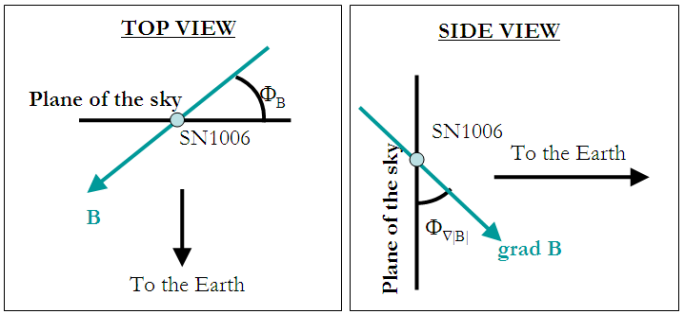}}
  \caption{{\em Left panel:} Definition of $\phi_{\bf B}$. 
   {\em Right panel:} Definition of $\phi_{\nabla |{\bf B}|}$.}

  \label{angles}
\end{figure}

Several contributions in the literature show that the observed morphology
of a SNR emitting synchrotron radiation is strongly affected by the
angle between the dominant direction of {\bf B} and the line
of sight, $\phi_{\bf B}$. Moreover, in our circumstance, since we
also have a gradient of the magnetic field, we are forced to consider
the dependence of the observed morphology on the angle between the
direction of the gradient and the plane of the sky, $\phi_{\nabla |{\bf
B}|}$. The definitions of $\phi_{\bf B}$ and $\phi_{\nabla |{\bf B}|}$
are sketched in Figure \ref{angles}. For the purposes of this work,
we have chosen to synthesize our maps in the following way: first,
we apply a rotation of $\phi_{\nabla |{\bf B}|}$ degrees around an axis
passing through the center of the remnant and parallel to the limbs,
where positive angles mean that regions of increasing {\bf B} are closer
to us. Next, we apply a rotation of $\phi_{\bf B}$ degrees around an
axis passing through the remnant center and lying in the plane of the
sky. Others rotation schemes give similar results. The adopted values
of $\phi_{\bf B}$ are from $0^\circ$ to $90^\circ$ in steps of $2^\circ$
and the ones of $\phi_{\nabla |{\bf B}|}$ are  from $0^\circ$ to $90^\circ$
in steps of $15^\circ$. Therefore, for each model, we have generated 315
maps encompassing all the combinations of the relevant angles.

\subsubsection{The morphological parameters}

\begin{figure}
  \centerline{\includegraphics[width=7.0cm]{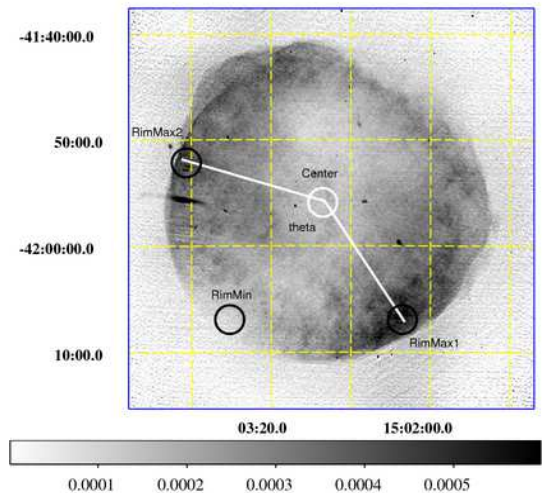}}
  \caption{Radio map of SN 1006 at 1.4 GHz, with the four regions in
  which the surface brightness of the remnant has been averaged for
  the purpose of computation of the four morphological parameters:
  $A=RIM_{max}/RIM_{min}$,
  $C=center/RIM_{max}$, $R_{max}=RIM_{max1}/RIM_{max2}$ and $\theta_D$, the aperture angle between the limb maxima. Adapted from \protect\citet{pdc09}.}

  \label{radiomap}
\end{figure}

The exploration of the parameter space is a challenging task, because
it involves the morphological comparison between many model maps and
the real SN 1006 images. We devised a simple procedure which involves
the comparison of the value of 4 morphological parameters calculated
both from the synthesized radio emission maps and from the observed
radio image of SN 1006. The parameters are the ratio between the
maximum and the minimum surface brightness ($S_b$) around the rim
($A$), the ratio between the maximum around the rim and the center of
the remnant ($C$), the ratio of $S_b$ of the two bright limbs
($R_{max}$), and the angular separation between the limbs
($\theta_D$). The parameters $A$, $R_{max}$ and $\theta_D$ were also
used in \citet{obr07}, and we refer to that paper for further
discussion. The $C$ parameter is introduced here to measure the
luminosity contrast between the brightest rim and the center. Note
that the $C$ parameter does not correspond to the $R_{\pi /3}$
parameter used by \citet{rbd04} to exclude the equatorial belt
scenario for SN 1006, because $C$ is measured in a small circular
region in the radio map (see below). The parameters $A$, $C$ and
$R_{max}$ are measured using the radio map of \citet{pdc09}, by
averaging the $S_b$ value in circular regions of 45\arcsec\ radius
(Fig. \ref{radiomap}). The values we obtained for the observed image
of SN 1006 are $A=2.7\pm0.2$, $C=0.36\pm0.03$, $R_{max}=1.2\pm0.1$, and
$\theta_D=135^\circ\pm10^\circ$. For the measurement of the parameters
in the synthesized model radio maps, we used an automated procedure to
find the maximum of the two limbs, the minimum between rims along the
rim and the central position. Then we used an average in a circular
region whose radius is the same percentage of the SNR radius as used
in the real radio image (5\%). Since the model images have been
synthesized using the same number of pixels per radius of the remnant,
this procedure ensures that the model and observed values of the
parameters are comparable.

\subsection{Results}

\subsubsection{Quasi-parallel scenario}

\begin{figure}
  \centerline{\includegraphics[width=9.0cm]{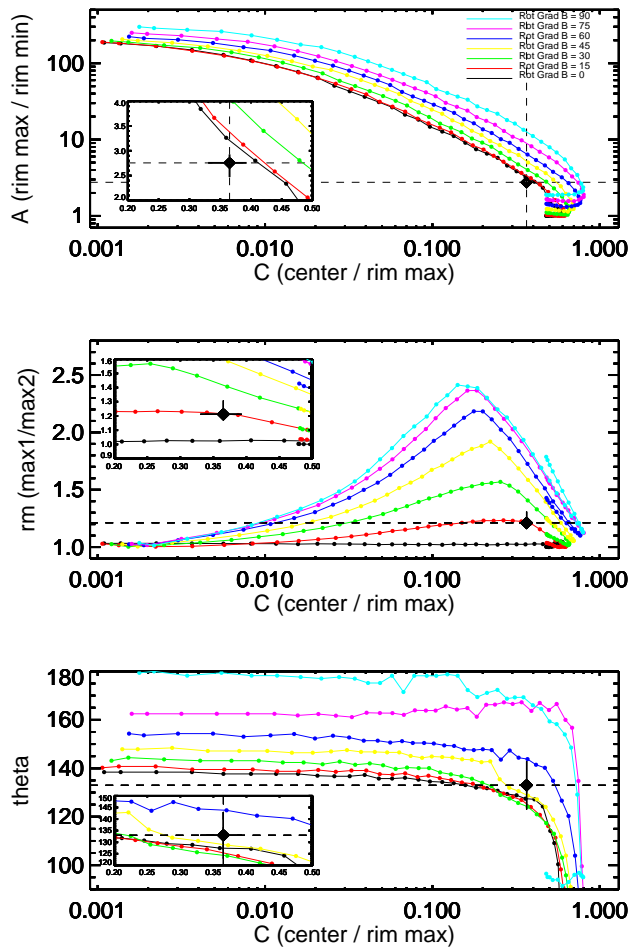}}
  \caption{{\em Top panel:} $A-C$ scatter plot for our models, assuming
the quasi-parallel scenario and an age of 1000 yr. Different colors
  correspond to different values of $\phi_{\nabla
|{\bf B}|}$, where each dot corresponds to a given value of $\phi_{\bf
B}$ (from $0^\circ$ to $90^\circ$ in steps of $2^\circ$. We overplotted
the $(A,C)$ values measured for SN 1006 (the inset shows a zoom near this
point). 
{\em Middle panel:} Same as top panel but for the $R_{max}-C$ pair.
{\em Lower panel:} Same as top panel but for the $\theta_D-C$ pair.}

  \label{resqpar}
\end{figure}

\begin{figure}
  \centerline{\includegraphics[width=9.0cm]{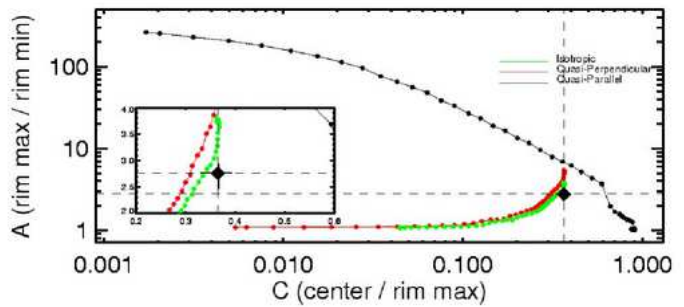}}
  \caption{Same as Fig. \protect\ref{resqpar}, top panel, but for the model 
   with uniform magnetic field.}
  \label{nograd}
\end{figure}

In Fig. \ref{resqpar}, we show the $A-C$, $R_{max}-C$ and $\theta_D-C$
scatter plots computed in the synthesized radio map of our SN 1006 model
including a weak gradient of the magnetic field ({\bf B} varies by a
factor of 1.4 over 10 pc), and considering a quasi-parallel scenario
for the obliquity dependence of the electron injection efficiency. We
also overplotted the values of the parameters derived using the real
radio map of the remnant of \citet{pdc09}. Though the agreement is not
exactly perfect in all the plots, we can note that we can define
a very limited region of the parameter space $\phi_{\bf B}-\phi_{\nabla
|{\bf B}|}$ which is compatible with the observed values of $A$, $C$,
$R_{max}$ and $\theta_D$. This means that the observed radio morphology
of SN 1006 is overall compatible with a quasi-parallel scenario for this
remnant, if we include a weak gradient of {\bf B}. A nonzero gradient
is an essential ingredient to reconcile the radio morphology with the
quasi-parallel scenario, and this is shown in Fig. \ref{nograd}, in
which we show the $A-C$ scatter plot for various $\phi_{\bf B}$ angles,
in the uniform {\bf B} case and quasi-parallel, quasi-perpendicular and
isotropic scenarios\footnote{In the uniform {\bf B} case, $R_{max}=1$
and $\theta_D=180^\circ$ always.}. In this case, this plot suggests that
quasi-parallel models do not reproduce the observed parameters at all,
unlike the quasi-perpendicular and isotropic scenarios, as already
pointed out by \citet{pdc09}.

\begin{figure*}
  \centerline{\hbox{
   \includegraphics[width=6.0cm]{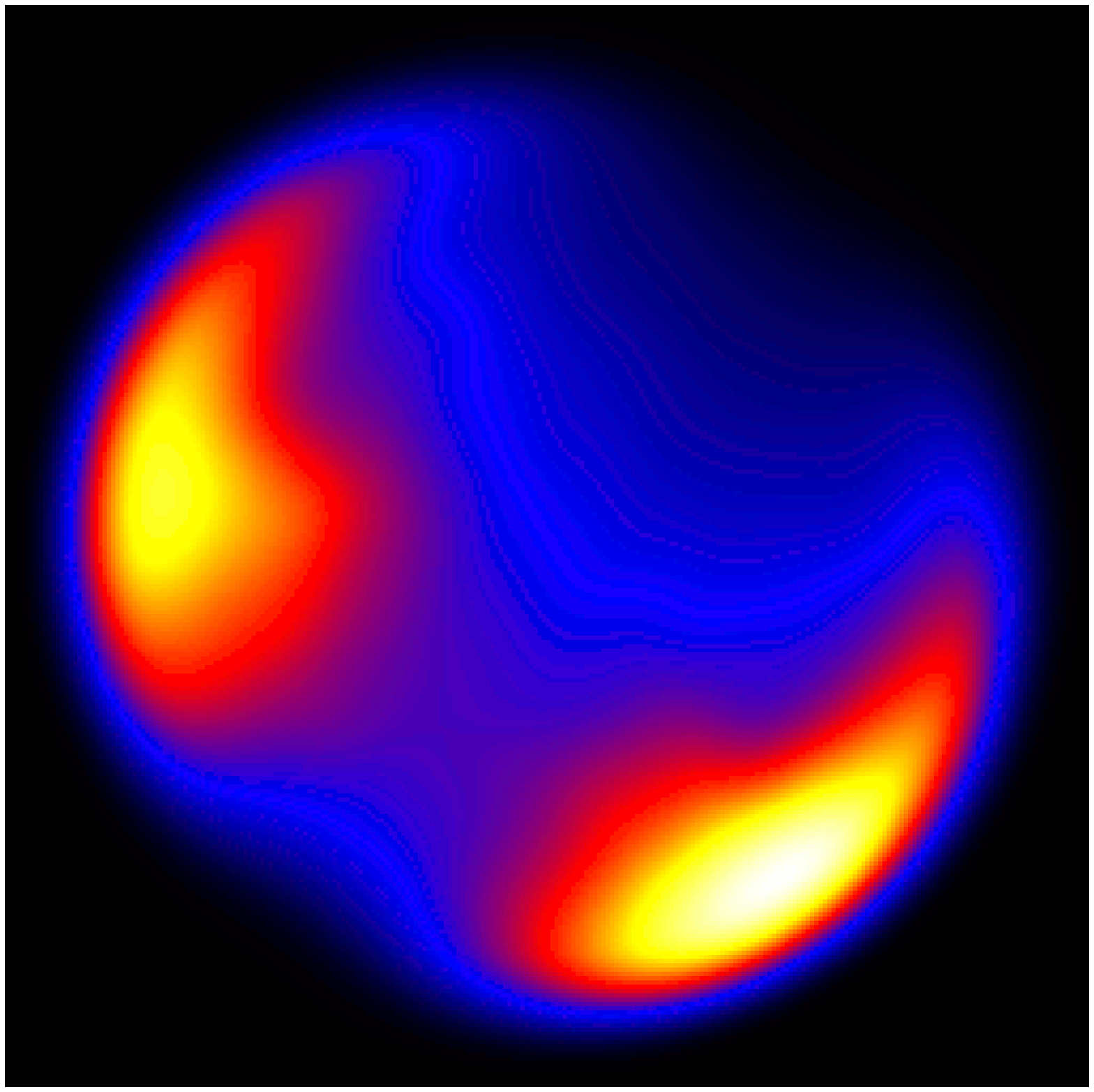}
    \includegraphics[width=6.0cm]{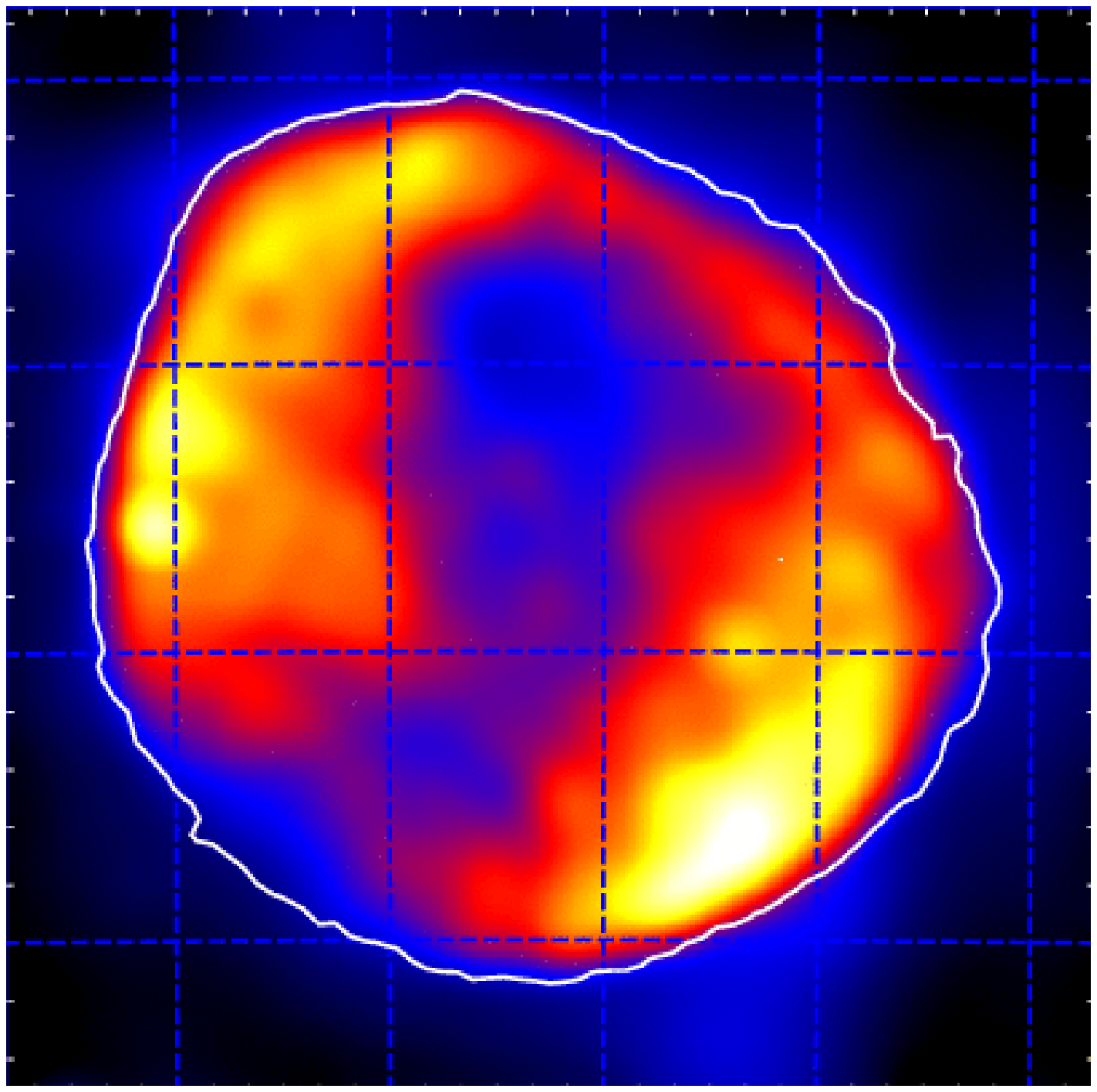}
    \includegraphics[width=6.0cm]{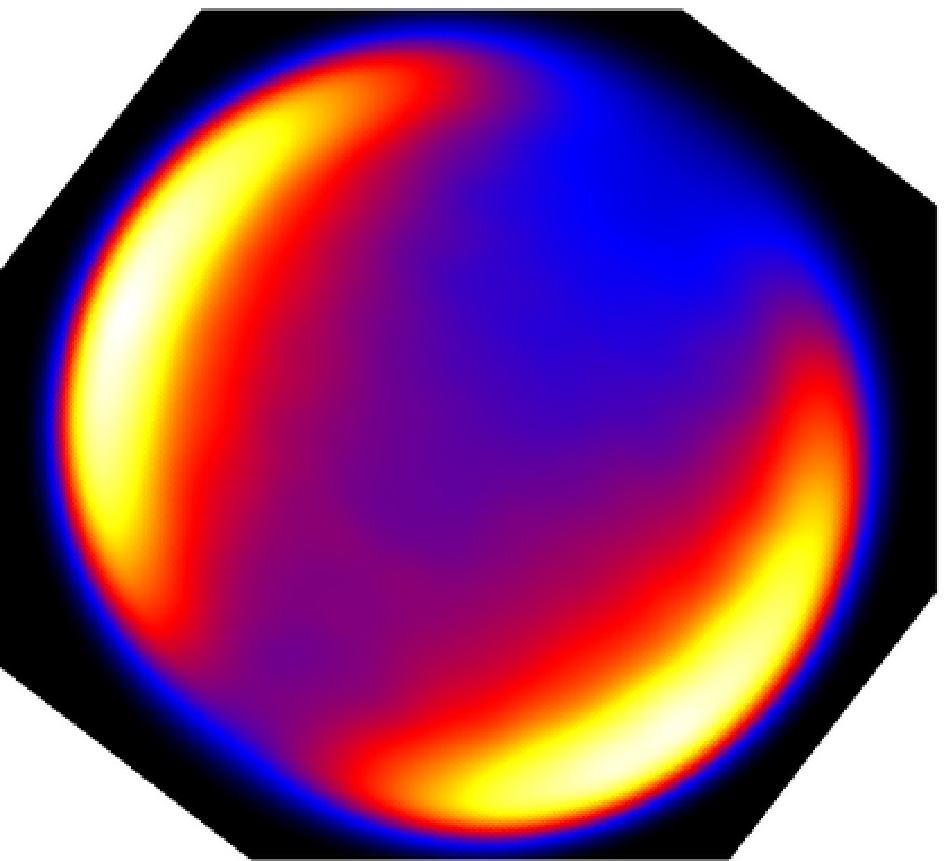}
    }}

  \caption{{\em Left panel:} Synthesized radio image at 1 GHz for our
best-fit model (GRAD1 in Table \protect\ref{models}, and
quasi-parallel injection efficiency scenario, $\phi_{\bf
B}=38^\circ\pm 4^\circ$ and $\phi_{\nabla |{\bf B}|}=15^\circ\pm
15^\circ$), smoothed with a sigma of $2^\prime$. {\em Central panel:}
radio map of SN 1006 of Fig. \ref{radiomap}, but smoothed with the same
sigma used for the model image. {\em Right panel:} same as the left
panel, but for the best-fit quasi-perpendicular scenario ($\phi_{\bf
B} \sim 38^\circ$).}

  \label{obsmod}
\end{figure*}

Let us now come back to Fig. \ref{resqpar}. A remarkable result is
that the comparison between the quasi-parallel model and the
observation strongly excludes a situation in which the polar caps are
in the plane of the sky ($\phi_{\bf B}=90^\circ$) or along the line of
sight ($\phi_{\bf B}=0^\circ$). The latter geometry would cause a
centrally brightened morphology instead of two limbs, as already
discussed by \citet{obr07}. The best-fit values of the aspect angle we
derive from Fig. \ref{resqpar} and a conservative estimate of their
uncertainties are $\phi_{\bf B}=38^\circ\pm 4^\circ$ and $\phi_{\nabla
|{\bf B}|}=15^\circ\pm 15^\circ$.

The synthesized radio map of the best fit model is shown in
Fig. \ref{obsmod}, along with the observed radio map. Both maps have
been smoothed to hide the small scale features and to focus the
comparison on large spatial scales. In fact, we don't expect a pixel
to pixel match between model and observations, given the simple
assumptions behind the model. Indeed, the large structures of the
observed radio emission are very well recovered by the best-fit model
and the two images look similar.

The comparison with other gradients, namely the GRAD2, GRAD3 and GRAD4
models, shows that it is not possible to find a satisfactory fit for
all the parameters. In particular, for higher values of the gradient,
the angle $\theta_D$ is underestimated, so GRAD1 is the only model
which gives us an overall good fit.

\subsubsection{Quasi-perpendicular scenario}

\begin{figure}
  \centerline{\includegraphics[width=9.0cm]{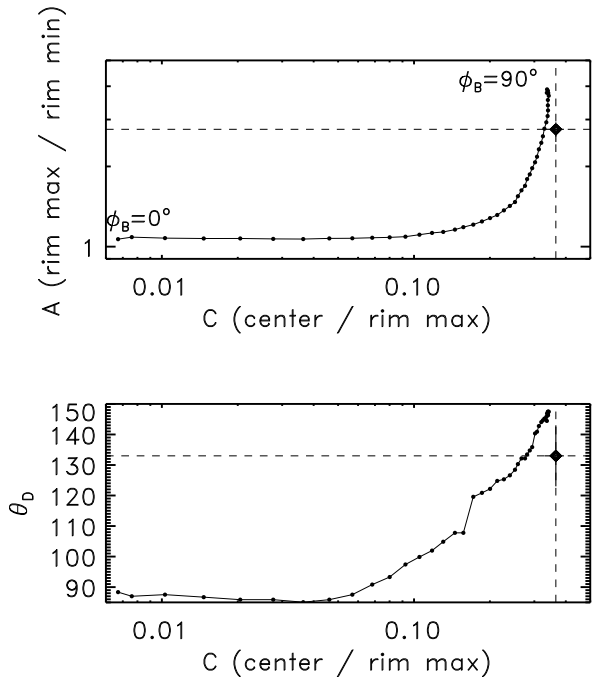}}
  \caption{{\em Top panel:} Same as Fig. \protect\ref{resqpar} (top panel)
  but for quasi-perpendicular scenario. There is no $\phigb$ angle in
  this case, because {\bf B} is always aligned with $\gb$. {\em Lower
  panel:} Scatter plot of $\theta_D-C$ parameters. We overplotted the
  values observed in SN 1006.}

  \label{resqper}
\end{figure}

We also produced model images in the quasi-perpendicular
injection scenario, and we repeated the analysis described in the
previous paragraphs. In this case, the direction of the $\gb$ is
aligned with the direction of {\bf B}, so the angle $\phigb$ is the
same as $\phib$, and we did not consider it any further. Another
consequence of this is that $R_{max}$ is always 1, so this parameter
cannot give any diagnostic.  The results are shown in
Fig. \ref{resqper}, and it seems that a good fit can be found for
$\phib \sim 70^\circ$ (a value in agreement with the azimuthal profile
analysis of \cite{pdc09}), even if the model points in the
$\theta_D-C$ diagram seems to be more distant from the observed points
than any scatter plot in Fig. \ref{resqpar}, thus indicating a better
fit in the quasi-parallel case than the quasi-perpendicular case.

\subsection{Discussion}

\begin{figure}
  \centerline{\includegraphics[width=9.0cm]{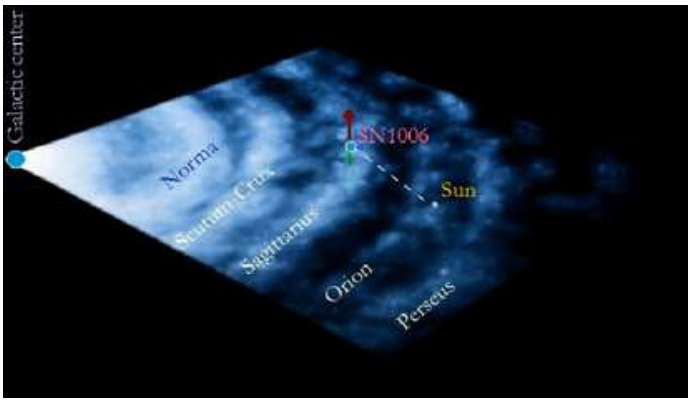}}
\caption{Artist's impression of the magnetic field at the location
of SN 1006 in our Galaxy. The red and green arrows mark the directions
of the field and its gradient as derived by the best-fit model for the
synchrotron radio emission of SN 1006, assuming a quasi-parallel
scenario, derived in this work. The figure is to scale.}

  \label{magf3d}
\end{figure}

The proposed method of comparison between models of synchrotron radio
emission and the real observational data of bilateral SNRs is based on
the calculation of 4 morphological parameters, and we have seen that
we get a good fit for a quasi-parallel injection
efficiency scenario if we include a gradient of {\bf B} and a worse
fit in case of quasi-perpendicular scenario with the same
gradient. This means that an agreement is found between the overall
observed morphology of SN 1006 and our model, which corresponds to well
defined values of the viewing geometry angles ($\phi_{\bf
B}=38^\circ\pm 4^\circ$ and $\phi_{\nabla |{\bf B}|}=15^\circ\pm
15^\circ$). We know that the angle between the SN 1006 axis of symmetry
and the Galactic plane is roughly $90^\circ$, and that its distance is
2.2 kpc and its Galactic latitude is $14.6^\circ$. Combining all this
information, we can plot the direction of the magnetic field and its
gradient in a 3D representation of the Galactic disk. This is shown in
Fig. \ref{magf3d}.

Remarkably, the direction of $\nabla |{\bf B}|$ points down toward the
plane and the direction of {\bf B} is aligned with the direction of
the spiral arm near the remnant. This is indeed a very reasonable
configuration for the magnetic field at this position in the Galaxy, because models of Galactic {\bf B} indicates that at this location above the plane the field still retaines spiral arms features, even if having smaller amplitudes.
Therefore, it is tempting to conclude that we are sampling the large scale
field of the Galaxy. However, we note that the best-fit value of
$\nabla |{\bf B}|$ (i.e. a variation of 1.4 times over 10 pc) seems to
be high to be associated with the large scale {\bf B} of the Galaxy. We
have used the model of \citet{srw08} to compute the expected variation
of the azimuthal field in the Galaxy at the location of SN 1006. In
particular, we have used their {\em ASS+RING} model which seems to be
favored by rotation measures of pulsars in the plane. At a
distance of $\sim 550$ pc from the plane for the SN 1006 remnant, the
large-scale field is expected to vary by a factor of 1.4 on scales of
more than 100 pc, much longer than the scale-length required by our
model fit. This opens up the possibility that we are actually sampling
the random magnetic field component. \citet{ms96} reports length
scales for this component of the order of few pc, which is compatible
with the variations we derive.

\subsection{Summary and conclusion}

The synchrotron radio emission of the archetypical bilateral supernova
remnant SN 1006 is compared against an MHD and emission model for this
remnant, including a gradient of the external magnetic field, particle
acceleration and its obliquity dependence, namely
quasi-perpendicular, quasi-parallel, and isotropic scenarios for the
injection efficiency. In order
to explore the parameter space, which is very large due to the
necessity of including the viewing geometry in the model-data
comparison, we developed a simplified procedure based on the
computation of four morphological parameters. We have found a very
good fit with a model assuming quasi-parallel injection efficiency and
an aspect angle $\phi_{\bf B}=38^\circ\pm 4^\circ$ between the
direction of {\bf B} the line of sight, and $\phi_{\nabla |{\bf
B}|}=15^\circ\pm 15^\circ$ between the plane of the sky and the
direction of the gradient of the magnetic field, and a variation of
{\bf B} of about 1.4 times over a scale of 10 pc. A worse fit is
obtained with quasi-perpendicular scenario. The overall morphology of
the observed radio emission at 1.4 GHz is correctly recovered by our
best-fit model. The projected direction of {\bf B} and $\nabla |{\bf
B}|$ in the Galaxy are along the spiral arm and toward the plane
respectively, which is in very good agreement with the expected
direction of the large scale Galactic {\bf B}. However, the implied
gradient is too high to be associated with the large-scale Galactic {\bf
B} and more typical of the random magnetic field components.

The application of our method to selected samples of bilateral
supernova remnants may yield independent estimates of the geometry of
the Galactic field at several locations, which can be useful to
understand the field topology in our Galaxy.

\begin{acknowledgements}
SPR acknowledges support from NASA and NSF for supernova-remnant
research.  FB thanks his collaborators S. Orlando, O. Petruk and
M. Miceli for their work on the modeling of the radio morphology of SN
1006.  BMG acknowledges the support of a Federation Fellowship from
the Australian Research Council.
\end{acknowledgements}

\end{document}